\magnification=1200 

\headline{\ifnum\pageno=1 \nopagenumbers 
\else \hss\number \pageno \fi} 
\def\sq{\hbox {\rlap{$\sqcap$}$\sqcup$}} 
\overfullrule=0pt
\font\boldgreek=cmmib10 
\textfont9=\boldgreek
\mathchardef\mymu="0916 
\footline={\hfil}
\font\boldgreek=cmmib10
\textfont9=\boldgreek
\mathchardef\myalpha="090B
\def\bfalpha{{\fam=9 \myalpha}\fam=1}  
\baselineskip=10pt 
\vbox to 1,5truecm{}
\parskip=0.2truecm 
\centerline{\bf FACTORIZATION OF SPANNING TREES}
\medskip
 \centerline{\bf ON FEYNMAN GRAPHS}\bigskip 
\centerline{\it (revised version)}
\bigskip \centerline{by}\smallskip 
\centerline{R. Hong Tuan}  
\medskip
 \centerline{Laboratoire de Physique Th\'eorique et Hautes Energies
\footnote{*}{Laboratoire associ\'e au Centre National de la Recherche
Scientifique - URA D0063}}  \centerline{Universit\'e de Paris XI, b\^atiment 
211, 91405
Orsay Cedex, France}  
\bigskip\bigskip
\baselineskip=20pt 
\noindent 
${\bf Abstract}$ \par 
In order to use the Gaussian representation for propagators
in Feynman amplitudes, a representation which is useful to relate string
theory and field theory, one has to prove first that each $\alpha$-parameter
(where $\alpha$ is the pa\-ra\-me\-ter associated to each propagator in the
$\alpha$-representation of the Feynman amplitudes) can be replaced by a 
constant
instead of being integrated over and second, prove that this constant can be 
taken
equal for all propagators of a given graph. The first proposition has been 
proven in
one recent letter when the number of propagators is infinite. Here we prove 
the
second one. In order to achieve this, we demonstrate that the sum over the 
weighted
spanning trees of a Feynman graph $G$ can be factorized for disjoint parts 
of $G$. The
same can also be done for cuts on $G$, resulting in a rigorous derivation of 
the
Gaussian representation for super-renormalizable scalar field theories. As a
by-product spanning trees on Feynman graphs can be used to define a 
discretized
functional space.

\vbox to 1cm{}

\noindent LPTHE Orsay 92/59 \par 
\noindent May 1995
\vfill\supereject \noindent {\bf 1. Introduction} \medskip In the study of
the relationship between field theories and string theories, the
$\alpha$-representation for Feynman graphs is a very useful tool [1-4]. In
this representation one $\alpha$-parameter is assigned to every propagator
and the only integrations to be made are over these parameters, integration
over the momenta circulating in the graph having already been made. This is
therefore a very economical representation and it has quite a while ago been
used to study the renormalization of field theories in the most accurate
way [5]. However, it has another nice property ; writing a propagator of a
scalar field theory as 
$$\lbrack (P_i - P_j)^2 \rbrack^{-1} = \int_0^{\infty} d \alpha \ exp \lbrack
- \alpha (P_i - P_j)^2 \rbrack \eqno(1)$$

\noindent we see that the $\alpha$-parameter is a sliding scale for
Gaussians. If we fix $\alpha$ to some constant $\bar \alpha$ we have what
we call a Gaussian representation (usually called a ``Gaussian
approximation") for the propagator. Now, discretized surface theories can be
constructed using precisely the Gaussian representation in planar graphs
amplitudes and taking $\bar \alpha$ to be the same quantity for {\it all}
propagators of a graph. $\bar \alpha$ is then interpreted as proportional to 
the
inverse of the slope of the Regge trajectories of the equivalent string 
theory [6].
In a recent letter [7] we proved that, indeed, once an integration over an 
overall
scale was made, all the $\alpha_i$'s ($\alpha_i$ being the $
\alpha$-parameter of
the propagator i) could be replaced by their mean-values $\bar \alpha_i$ 
which, in
turn, were demonstrated to be O(1/$I$), $I$ being the total number of 
propagators
of any one-particle, one vertex irreducible Feynman graph $G$, planar or
non-planar, $I \to \infty$. This was done for any number of Euclidean 
dimensions
where the theory was super-renormalizable. (When the theory is 
renormalizable we
have to make the weak assumption that a logarithm coming from the
renormalization of some sub-divergence is provoking only a shift in the
coupling constant when the integration is made over the momenta of the legs
of the sub-diverging part). However, in any case, there is a second step in
the derivation which consists in proving that all the $\bar \alpha_i$'s can,
in turn, be replaced by a single value $\bar \alpha$ for a given  graph $G$. 
This
could be demonstrated [7] provided that the sum over spanning trees of $G$
can be considered as a functional integral, i.e. that the {\it sum could be
factorized} on disjoint domains of $G$. It is our purpose here to present a
rigorous derivation of that statement and thereby ending the proof about the
validity of the Gaussian representation (using an unique $\bar \alpha$).\par 

In section 2 we present the basics of the $\bar{\alpha}$-representation for
Euclidean scalar field theories. We give the general expression for the 
Feynman
graph amplitude $F_G$ of a graph $G$ with $I$ internal lines, $L$ loops in 
terms
of an integral over $I$ $\alpha$-parameters. This integral can be evaluated
using the mean-value theorem which states that if a function $f$ is 
continuous
in its arguments $\{ \alpha \}$, then  
$$\int_{\omega} f = V \ f(\{ \bar{\alpha} \}) \ \ ,$$

\noindent where $V$ is the volume of the connected domain $\omega$ over which
the integration extends and $\{ \bar{\alpha} \}$ a set of values of the
$\alpha$-parameters defining some point in $\omega$. We expect the 
mean-values
$\bar{\alpha}_i$ of the $I$ $\alpha$-parameters to be a priori different. 
Then,
the central result of this article is to demonstrate that all
$\bar{\alpha}_i$'s can be replaced by one {\it single} value $\bar{\alpha}$
without changing the value of Feynman amplitude (expressed as a function of 
the
$\bar{\alpha}_i$'s via the mean-value theorem). This will amount to
showing that the ratio of polynomials $Q_G(P_v, \{ \bar{\alpha}_i \} )$
defined in section 2 and appearing in the expression of $F_G$ is indeed
insensitive to that replacement. Isolating one particular
$\bar{\alpha}_i$, $Q_G$ can be set in the form 
$$Q_G ( P, \{ \bar{\alpha}_i \}) = (\bar{e}_i / \bar{b}_i) (\bar{d}_i / 
\bar{e}_i +
\bar{\alpha}_i) / (\bar{a}_i / \bar{b}_i + \bar{\alpha}_i)$$

\noindent where $\bar a_i / \bar{b}_i$ and $\bar{d}_i / \bar{e}_i$ are ratios
of homogeneous polynomials in the mean-values of all $\alpha$-parameters
except $\alpha_i$. The proof of the independence on the shift $\bar{\alpha}_i
\to \bar{\alpha}$ will then translate into a proof of the equality of the
ratios $\bar{a}_i / \bar{b}_i$ and $\bar{d}_i / \bar{e}_i$. Then, an 
important
property of $\bar{a}_i$, $\bar{b}_i$, $\bar{d}_i$ and $\bar{e}_i$ is that 
they
can be expressed as sums over products of $\bar{\alpha}_l$'s, $l$ indexing
propagators belonging to spanning trees of $G$ (a spanning tree of $G$ is a
tree incident with all vertices of $G$). The ratio $\bar{a}_i / \bar{b}_i$
involves a sum over trees containing $i$, i.e. $\bar{a}_i$, and a sum over
trees not containing $i$, i.e. $\bar{b}_i$. The ratio $\bar{d}_i / \bar{e}_i$
is the ratio of a sum over trees containing $i$ and cut at some other
propagator over a sum over trees not containing $i$ and cut at some other
propagator (cutting means that the propagator is deleted from the tree,
yielding a cut). Then, proving that $\bar{a}_i / \bar{b}_i = \bar{d}_i /
\bar{e}_i$ amounts to proving that {\it the effect of cutting of a
propagator can be factorized in the sum over trees}. This will be true if
the structure of trees is such that their structure far from $i$ is
independent from their structure close to $i$. The next sections will be
devoted to the proof that the sums over spanning trees can indeed be
factorized over domains far apart on $G$.\par

In section 3 we restrict ourselves to the case of self-avoiding paths on
$G$ instead of trees. This is because, aside from simplicity in a first
approach, there is always a self-avoiding path linking two vertices of $G$
on any spanning tree of $G$. Thus, spanning trees can be built out of
self-avoiding paths. We first give the general strategy for the proof of the
equality $\bar{a}_i/\bar{b}_i = \bar{d}_i/\bar{e}_i$. Then, the following 
ratio
$$R_i(s_j) = \sum_P P(i, s_j) / \sum_P P(\bar{i}, s_j)$$
 
\noindent is proven to be independent of $s_j$ if $s_j$ is a vertex 
infinitely far
from $v_i$ on $G$, $v_i$ being a vertex incident with the propagator $i$. 
(The notion
of distance on $G$ will be discussed later on. By infinitely far we mean 
that an
infinite number of propagators separate $v_i$ from $s_j$). $P(i, s_j)$ is a
self-avoiding path linking $v_i$ to $s_j$ {\it going through i}. $P(\bar{i}, 
s_j)$ is
also a self-avoiding path linking $v_i$ to $s_j$ {\it but not going through 
i}. The
proof uses the evaluation of $R_i(s_j)$ as a mean-value in a volume $V_j$. 
In fact
$R_i(s_{j+1})$ will turn out to be the average of $R_i(s_j)$. Letting $j \to 
\infty$,
the averaging process, repeated an infinite number of times, removes the 
dependence
on $s_j$ of $R_i(s_j)$. This proof is essential as the same proof will be 
used to
treat $m$-paths, i.e. paths with $m$ connected parts, of which at least one 
of them
is a path $P(i, s_j)$ or a path $P(\bar{i}, s_j)$. We also discuss some 
possible
difficulties associated with the convergence of the averaging process. \par 

In section 4, the main difficulty which could impede convergence is 
identified with
the fact that the ratio $R_i(s_{j+1})/R_i(s_j)_M$ can be infinitesimally 
close
to one, $R_i(s_j)_M$ being an extremum value of $R_i(s_j)$ when the weight
for $R_i(s_j)_M$ is infinite with respect to the sum of the weights for all
other $R_i(s_j)$. \par
We solve this difficulty in the case of spanning trees in $V_j$ directly. 
Then,
$s_j$ is replaced by $\{s_j\}$, a partition of the vertices of the border of
$V_j$ with $V_{j+1} - V_j$, where $V_j$ is a volume which increases with
$j$. In the averaging procedure $W_m^{m'}(\{s_j \}, \{s_{j+1}\})$ is the
weight of $R_i^m(\{s_j\})$ in the evaluation of $R_i^{m'}(\{s_{j+1}\})$. The
weight-ratio 
$$\sum_{\{s_j \}_M} W_{m_M}^{m'} \left ( \{s_j\}_M , \{s_{j+1}\}
\right ) / \sum_{\{s_j \}} W_{m}^{m'} \left ( \{s_j\} , \{s_{j+1}\} \right 
)$$

\noindent is studied, where $\{s_j \}_M$ is a partition corresponding to
$R_i^m(\{s_j \}_M)$, i.e. an extremum value of $R_i^m(\{s_j \})$ and $\{s_j 
\}$
any other partition. When the above weight-ratio becomes infinite we have a 
convergence
problem in the case of spanning trees equivalent to the one for paths 
mentioned
above. We then prove that if a constraint on the construction of the $V_j$'s 
is
imposed which, indeed, is easy to implement, the above weight-ratio takes 
the same
value for all partitions $\{s_{j+1}\}$ of the vertices on the border of 
$V_{j+1}$
with $G - V_{j+1}$. \par

This result allows to conclude that if the above weight-ratio is infinite, 
it is
infinite whatever $\{s_{j+1}\}$ and then $R_i^{m'}(\{s_{j+1}\})$ is equal to
$R_i^m(\{s_j\}_M)$, implying that convergence has been obtained. When,
this weight-ratio is finite, then the convergence of the averaging process 
is not
impeded and a unique value for $R_i^{m'}(\{s_{j+1}\})$ is obtained as $j \to
\infty$. \par  

In the last sub-section we show how the averaging process works in the case 
of
spanning trees on $G$ (instead of paths or multiple paths).\par  

In section 5, the use of the above proof allow the proof of the 
factorization theorem.
Section 6 will be the conclusion. \par  \vfill \supereject

\noindent {\bf 2. The $\bfalpha$-representation} \smallskip 

Here, we deal with scalar field theories in $d$ Euclidean dimensions. We 
study
one-line, one-vertex irreducible Feynman graphs with $I$ internal lines
(propagators), $L$ loops, external momenta $P_v$ and we take the coupling 
cons\-tant
equal to - 1 in order to simplify. Then, $F_G$, the Feynman amplitude for a 
graph $G$
and for a field of mass $m$ reads [8] 
$$F_G = h_0 (4 \pi)^{-dL/2} \int_0^{h_0} \ \lbrack \prod_{i=1}^I d \alpha_i 
\rbrack \
\delta (h_0 - \sum_i \alpha_i) \ \lbrack P_G(\alpha) \rbrack^{-d/2} .$$
$$\int_0^{\infty} d \lambda / \lambda \ \lambda^{I-dL/2} \ exp \lbrace - 
\lambda
\lbrack Q_G(P_v, \alpha) + m^2 h_0 \rbrack \rbrace \eqno(2)$$ 

\noindent where $P_G(\alpha)$ is a homogeneous polynomial of degree $L$ in 
the
$\alpha_i$'s defined as 
$$P_G(\alpha) = \sum_{\cal T} \prod_{l \notin \cal T} \alpha_l \eqno(3)$$ 

\noindent where the sum runs over all the spanning tree ${\cal
T}$ of $G$. (A spanning tree of $G$ is a tree incident with every vertex of 
$G$).
$Q_G(P, \alpha)$ is quadratic in the $P_v$'s and is given by the ratio of a
homogeneous polynomial of degree $L$+1 over $P_G(\alpha)$ 
$$Q_G(P_v, \alpha) = \lbrack P_G(\alpha) \rbrack^{-1} \sum_{\cal C} s_{\cal 
C}
\prod_{l \in \cal C} \alpha_l \eqno(4)$$

\noindent where the sum runs over all cuts \ ${\cal C}$ of $L$+1 lines that
divide $G$ in two connected parts $G_1({\cal C})$ and $G_2({\cal C})$, with
$$s_{\cal C} = (\sum_{v \in G_2({\cal C})} P_v)^2 = (\sum_{v \in G_1({\cal
C})} P_v)^2 \eqno(5)$$

\noindent (A cut ${\cal C}$ is obtained from a tree ${\cal T}$ by cutting off
one line of ${\cal T}$. Then, the cut ${\cal C}$ will consist of all lines on
$G$ not on ${\cal T}$ {\it plus} the line of $\cal{T}$ which has been cut).
We note that $\lambda$ can be interpreted as an overall scale for the
$\alpha$-parameters ((2) indicates that $\displaystyle {\sum_i} \alpha_i = 
h_0$
where $h_0$ is arbitrary but taken equal to one in most circumstances). The
integration over $\lambda$ gives the overall divergence of $F_G$ for a
renormalizable theory. Here, we will limit ourselves to super-renormalizable
theories, i.e. $I-dL$/2 will always be positive, giving a convergent integral
\vfill \supereject
$$I_{\lambda} (Q_G) = \int_0^{\infty} d \lambda /\lambda \ \ \lambda^{I-dL/2}
exp \lbrace - \lambda \lbrack Q_G(P_v, \alpha) + m^2 h_0 \rbrack \rbrace$$ 
$$=
\Gamma (I - dL/2) \ \lbrack Q_G(P_v, \alpha) + m^2 h_0 \rbrack^{-(I-dL/2)}
\eqno(6)$$
 
Now, the spirit of the demonstration concerning the replacement of the
$\alpha_i$'s by their mean-values $\bar \alpha_i$ consists in isolating the
dependence of the integrand of $F_G$ on one particular $\alpha_i$ and in 
using
the mean-value theorem to perform the integration [7]. A discussion of the
consistency of the result of this integration then shows that one should
have, in any case [7], 
$$\bar \alpha_i = O(h_0/I) \eqno(7)$$

\noindent This property can easily be understood by considering the phase 
space
for the $I$ variables $\alpha_i$, which can be found to be equal to 
$h_0^{I-1}/(I -
1)! \sim (e \ h_0/I)^I$, leaving a phase space for each $\alpha_i$ of the 
order of
$e \ h_0/I$. \par
 
Then, one has to show that indeed all $\bar \alpha_i$'s can be
taken equal to some common value $\bar \alpha$. Consequently, we shall {\it
define}  $\bar \alpha$ by 
$${\cal N}_{\cal T} \bar \alpha^L = \sum_{\cal T} \prod_{l \notin \cal T} 
\bar
\alpha_l = P_G(\bar \alpha) \eqno(8)$$

\noindent where ${\cal N}_{\cal T}$ is the number of spanning trees on $G$. 
We
see from (2) that the expression obtained for $F_G$ by using the mean  value
theorem is 
$$F_G = h_0 (4 \pi)^{-dL/2} \lbrack P_G(\bar \alpha) \rbrack^{-d/2}
I_{\lambda} \lbrack Q_G(P_v, \lbrace \bar \alpha_i \rbrace) \rbrack \
h_0^{I-1}/(I -1) ! \eqno(9)$$

\noindent where the factor $h_0^{I-1}/(I - 1) !$ is the volume of the phase
space available for the $\alpha_i$'s. So, in fact, from (9) it is clear that 
our
goal amounts to showing that $Q_G(P_v, \lbrace \bar \alpha_i \rbrace )$ is 
not
affected by the replacement $\bar \alpha_i \to \bar \alpha$. Let us write
$$P_G(\bar \alpha) = \bar a_i + \bar b_i \bar \alpha_i \eqno(10)$$
$$\sum_{\cal C} s_{\cal C} \prod_{l \in \cal C} \bar \alpha_l = \bar d_i +
\bar e_i \bar \alpha_i \eqno(11)$$

\noindent where $\bar a_i, \bar b_i, \bar d_i$ and $\bar e_i$ do not contain
any $\bar \alpha_i$ factor. As 
$$Q_G (P_v, \lbrace \bar \alpha_i \rbrace ) = (\bar
e_i/ \bar b_i) (\bar d_i/ \bar e_i + \bar \alpha_i)/(\bar a_i/ \bar b_i + 
\bar
\alpha_i) \eqno(12)$$

\noindent it is easily seen that if 
$$\bar d_i/\bar e_i = \bar a_i/\bar b_i \ \ \ , \eqno(13)$$

\noindent the shift $\bar \alpha_i \to \bar \alpha$ will not affect
$Q_G(P_v, \lbrace \bar \alpha_i \rbrace )$. Repeating the reasoning for all $
\bar
\alpha_i$'s shows that $Q_G(P_v, \lbrace \bar \alpha_i \rbrace )$ is 
invariant
under the replacement of all the $\bar \alpha_i$'s by $\bar \alpha$ if (13) 
is
true. It will be the purpose of the next sections to demonstrate that, up to
vanishing corrections as $I \to \infty$, (13) is indeed true. Then, the 
shift of Log
$[Q_G(P_v, \lbrace \bar \alpha_i \rbrace )]$ 
$$\delta Q_G(P_v, \lbrace \bar \alpha_i
\rbrace )/Q_G (P_v , \lbrace \bar{\alpha}_i \rbrace ) = Q_G^{-1} \sum_{i=1}^I
{\partial Q_G \over \partial \bar \alpha_i} \delta \alpha_i \eqno(14)$$

\noindent will be vanishing, because $\delta \alpha_i \sim 1/I$ for
any i $(\delta \alpha_i = \bar \alpha - \bar \alpha_i)$. Then, defining
$\Delta (\bar \alpha)$ as 
$$\Delta (\bar \alpha) = \prod_{l \in G} \bar \alpha_l \eqno(15)$$

\noindent i.e. defined as the product of all $\bar \alpha_l$'s over $G$, we 
can
write (see (3) and (10)) 
$$\bar a_i = \Delta (\bar \alpha) \sum_{ {\cal T} \supset i} \ \prod_{l \in 
\cal T}
\bar \alpha_l^{-1} \eqno(16)$$

$$\bar b_i = \bar \alpha_i^{-1} \Delta (\bar \alpha) \sum_{ {\cal T}
\not\supset i} \ \prod_{l \in {\cal T} } \bar \alpha_l^{-1} \eqno(17)$$

\noindent In an analogous way $\bar d_i$ and $\bar e_i$ can be written
$$\bar d_i = \Delta (\bar \alpha) \sum_{ {\cal T} \supset i} \ \prod_{l \in
{\cal T} } \bar \alpha_l^{-1} \sum_{k \in {\cal T}\atop k \not= i} \bar
\alpha_k \ s_{{\cal C}_k }\ \nu^{-1}({\cal C}_k) \eqno(18)$$

$$\bar e_i = \bar \alpha_i^{-1} \Delta (\bar \alpha) \sum_{ {\cal T}
\not\supset i} \ \prod_{l \in \cal T} \bar \alpha_l^{-1} \sum_{k \in {\cal T}
} \bar \alpha_k \ s_{ {\cal C}_k} \ \nu^{-1}({\cal C}_k) \eqno(19)$$

\noindent where $\nu ({\cal C}_k)$ counts the number of times the same cut
${\cal C}_k$ is obtained in cutting trees ${\cal T}$ at $k$, $k$ being among 
the
$\nu ({\cal C}_k)$ propagators binding the connected parts of $G$, $G_1({\cal
C}_k)$ and $G_2({\cal C}_k)$ separated by ${\cal C}_k$. \par 

To understand how (18) and (19) can be obtained from (11), let us recall 
that we are
summing in (11) over all possible cuts $\cal{C}$ belonging to $G$. It is 
then, useful
to note that the same cut ${\cal C}_{k}$, containing the propagator $k$, can 
be
obtained by cutting different trees provided these trees have exactly the 
same
structure in $G_1({\cal C}_{k})$ and $G_2({\cal C}_{k})$ and only in this 
case.
That is, they will only differ by the propagator on them linking
$G_1({\cal C}_{k})$ and $G_2({\cal C}_{k})$. Let us note by $\nu({\cal 
C}_{k})$
the number of propagators on $G$ linking $G_1({\cal C}_{k})$ and
$G_2({\cal C}_{k})$. Then, $\nu({\cal C}_{k})$ will count how many trees
$\cal{T}$ can be cut to yield the same cut ${\cal C}_{k}$. Dividing by
$\nu({\cal C}_{k})$ in (18) and (19) ensures that each cut is only counted 
once
when cutting all possible trees yielding it. ($s_{{\cal C}_{k}}$ is defined 
by
(5) where ${\cal C}$ is ${\cal C}_{k}$). Of course, all possible cuts are
generated because $k$ is taken to be any propagator of $G$ on $\cal{T}$. \par

Comparing (18) and (19) to (16) and (17) respectively, we see that the sum
\break
\noindent $\displaystyle {\sum_{k \in \cal T} }  \bar \alpha_k \ s_{{\cal 
C}_k} \
\nu^{-1}({\cal C}_k)$ is the factor which distinguishes $(\bar a_i, \bar 
b_i)$
from $(\bar d_i, \bar e_i)$. If this factor can be  factorized out of the sum
over trees, the relation (13) will be obvious. However, this can only be 
done if
the structure of the trees far from $i$ is independent of their structure 
near
$i$ and when, in addition, $k$ and $i$ are far apart on $G$, i.e. if they are
separated by an infinite number of propagators on $G$. When $I \to \infty$, 
most
of the propagators of $G$ will be far from $i$, so that we will be able to
neglect in the sum over $k$, those $k$ which are in a finite range of $i$. So
our main goal will be, in fact, to show that {\it a factorization  occurs in 
the
sum of weighted trees for domains far apart on} $G$.\par  \vfill \supereject 

\noindent {\bf 3. Construction of the spanning trees on G : paths on G} 
\medskip 

It will prove to be convenient for the following to rewrite (18) and (19) by 
inverting
the summation order, ${\cal T}_k^i({\cal C}_k)$ being a spanning tree which 
contains
$i$ and $k$ and which cut at $k$ gives a cut ${\cal C}_k$, containing $k$, 
$$\bar d_i = \Delta (\bar \alpha) \sum_k \bar \alpha_k \sum_{{\cal T}_k^i ({
\cal
C}_k)} s_{{\cal C}_k} \ \nu^{-1} ({\cal C}_k) \prod_{l \in {\cal T}_k^i} \bar
\alpha_l^{-1} \eqno(20)$$

$$\bar e_i = \bar \alpha_i^{-1} \Delta (\bar \alpha) \sum_k \bar \alpha_k
\sum_{{\cal T}_k ({\cal C}_k^i)} s_{{\cal C}_k^i} \ \nu^{-1} ({\cal C}_k^i)
\prod_{l \in {\cal T}_k} \bar \alpha_l^{-1} \eqno (21)$$

\noindent where ${\cal T}_k({\cal C}_k^i)$ contains $k$ but ${\it not}$
$i$ and gives a cut ${\cal C}_k^i$ when cut at $k$, ${\cal C}_k^i$ 
containing $i$ and
$k$. What we want to demonstrate now is that {\it for any $k$ far
apart from $i$ on $G$}, defining 
$$\bar d_{i,k} = \sum_{{\cal T}_k^i ({\cal C}_k)} s_{{\cal C}_k} \ \nu^{-1}({
\cal
C}_k) \prod_{l \in {\cal T}_k^i} \bar \alpha_l^{-1} \eqno(22)$$

$$\bar e_{i,k} = \bar \alpha_i^{-1} \sum_{ {\cal T}_k ({\cal C}_k^i)} \ s_{
{\cal C}_k^i} \ \nu^{-1} ({\cal C}_k^i) \prod_{l \in {\cal T}_k} \bar
\alpha_l^{-1} \eqno(23)$$

\noindent we have 
$$\bar d_{i,k}/ \bar e_{i,k} = \bar a_i / \bar b_i \eqno(24)$$

\noindent up to terms which tend to zero as $I \to \infty$. The proof of 
(24) naturally
entails the validity of (13) because $\bar d_i = \Delta (\bar \alpha) 
\displaystyle{\sum_k} \bar \alpha_k \ \bar d_{i,k}$ and $\bar e_i = \Delta
(\bar \alpha) \displaystyle{\sum_k} \bar \alpha_k \ \bar e_{i,k}$ and also
because those $k$ within a ``volume'' $V_j$ containing $i$ and a number of 
propagators
infinitesimal compared to the total number in $G$ contribute to a negligible 
fraction of the
sum, as will be shown at the end of section 5. \par 

\medskip \noindent {\bf A - {\it General strategy for the proof of (24)}}\par

We now give the general lines of the proof of the relation (24). \par

In the first place we consider $G$ as embedded in a $R^3$-space embedded 
with a
metric. Then, we consider a volume $V_j$ in that space which contains the
propagator $i$. When $j$ is finite the number of propagators contained in 
$V_j$
will be finite. As $j \to \infty$ the number of propagators contained in 
$V_j$ will
tend to infinity. We will however consider the number of propagators on $G$ 
outside
$V_j$ infinite with respect to the number of those inside $V_j$, even as $j 
\to
\infty$. The propagator $k$ is taken to be outside $V_j$. Of course, in the 
sum over
all $k$'s some of them are inside $V_j$, but their number will be 
infinitesimal with
respect to the total number of $k$'s, i.e. the number of propagators in $G$. 
So the
contribution of $k$'s inside $V_j$ will be negligible in the sum over them. 
\par

The reason we want to isolate $V_j$ is that we shall see that inside $V_j$ 
we can
sum over sub-trees in it (when a spanning tree on $G$ is cut by the
border of $V_j$, the portion of that spanning tree inside $V_j$ has no
reason to be connected and is in general composed of several connected
pieces that we call sub-trees) independently of the rest of the trees 
outside it
provided the vertices on the border of $V_j$ are partitioned in a definite 
way, each
partition corresponding to a partition of sub-trees in $V_j$. So we get a {
\it
factorization} in the structure of trees on $G$, the sum over all trees in 
$G$ being
factorized into the sum over sub-trees in $V_j$ times the sum over all 
sub-trees in
$G - V_j$ {\it for a given partition of the border of $V_j$ in sub-sets of 
vertices},
each sub-set being attached to the one sub-tree in $V_j$. Of course, 
sub-trees in $G
- V_j$ (i. e. $G$ minus all propagators and vertices in $V_j$) have to be 
compatible
with those in $V_j$ (not forming loops for instance) in that one should 
obtain trees
in $G$ altogether. However, again this implies only a restriction of the 
partition of
vertices on the border of $V_j$, with respect to sub-trees in $G - V_j$ this 
time. So
given the structure on the border of $V_j$, factorization of trees inside 
$V_j$ and
outside it holds. \par 

However, we still have a dependence on the structure of the
partition of the border of $V_j$. That is where comes a first and essential 
part of
the proof. \par 

Let us call $T(i, \lbrace s_j \rbrace)$ those sets of sub-trees in
$V_j$ which contain the propagator $i$ and $T(\bar{i}, \lbrace s_j \rbrace)$ 
those
sets of sub-trees in $V_j$ which do not contain the propagator $i$, both 
being
attached to a partition $\lbrace s_j \rbrace$ of the border of $V_j$. Let us 
define
the ratio 

$$R_i(\{s_j\}) = \sum_T T(i, \{s_j \})/\sum_T T(\bar{i} , \{ s_j \})
\eqno(25)$$

\noindent which is the ratio of the sum of the weights of sub-trees sets 
$T(i,
\{s_j \})$ over the sum of the weights of sub-trees sets $T(\bar{i}, \{s_j 
\})$.
(Each propagator $\ell$ on a sub-tree brings a factor $\bar{\alpha}_{
\ell}^{-1}$
in the weight of any sub-tree). Then, it will be proved in the next 
sub-section
(for self-avoiding paths) and in the next section (for trees themselves) 
that as
$j \to \infty$, i.e. when the radius of $V_j$ grows to infinity, but with 
$V_j$
still being infinitesimal with respect to $G$, $R_i(\{s_j \})$ tends to some
value $R_i^{\infty}$ independent of the partition $\{s_j \}$. Using the
factorization property discussed above, this amounts to say that
$$\bar{a}_i/\bar{b}_i = R_i^{\infty} \bar{\alpha}_i \eqno(26.a)$$

\noindent (see (16) and (17)). Now, the same argument can be used to derive
also
$$\bar{d}_{i,k}/ \bar{e}_{i,k} = R_i^{\infty} \bar{\alpha}_i \eqno(26.b)$$

\noindent and thereby prove (24) if the additional structure of cutting
throught $k$ does not interfere with the inside of $V_j$, i.e. if $\nu({\cal
C}_k)$ and $s_{{\cal C}_k}$ are unaffected by the inside of $V_j$. \par

Let us denote by ${\cal S}_k$ the surface defined by the cut ${\cal C}_k$ 
going
through $k$ and dividing $G$ into two separate pieces $G_1({\cal C}_k)$ and
$G_2({\cal C}_k)$. (${\cal S}_k$ cuts through all the propagators on $G$ 
linking
$G_1({\cal C}_k)$ and $G_2({\cal C}_k)$). If ${\cal S}_k$ does not go 
through $V_j$,
the factorization property then trivially shows that the sums inside $V_j$ 
are
unaffected by ${\cal S}_k$, there is no interference. If ${\cal S}_k$ goes 
through
$V_j$, $\nu({\cal C}_k)$ counting the propagators in ${\cal S}_k$ will only 
be
infinitesimally affected by the structure of sub-trees inside $V_j$, as well 
as
$s_{{\cal C}_k}$, because the number of propagators in $V_j$ is 
infinitesimal with
respect with their total number in $G$. So there is only an infinitesimal
interference in this case and therefore the relations above prove the 
validity of
(24). \par 

In the next sub-sections, we simplify in a first approach, replacing trees
on $G$ by self-avoiding paths and prove in this case that the ratios
$R_i(\{s_{j+1} \})$ are indeed averages of $R_i(\{ s_j \})$. We also show
how the factorization described above works. The averaging property
means that
$$R_i(\{s_j \})_{min} \leq R_i(\{ s_{j+1} \}) \leq R_i(\{s_j \})_{max} 
\eqno(27)$$

\noindent where $R_i(\{s_j \})_{min}$ and $R_i(\{ s_j \})_{max}$ are
respectively the maximum and the minimum value of $R_i(\{ s_j \})$. If, in
the averaging, $R_i(\{ s_j \})_{min}$ or $R_i(\{ s_j \})_{max}$ only have
a {\it finite weight relative to the sum of the weights of the other
values of $R_i(\{ s_j \})$, then it is clear that $R_i(\{ s_{j+1} \})$
will be different from $R_i(\{ s_j \})_{min}$ or $R_i(\{ s_j \})_{max}$},
even having at least a {\it finite} (non-infinitesimal) difference with
them. Then, we will have 
$$R_i(\{ s_{j+1} \})/R_i (\{ s_j \} )_{max} = 1 - \eta_1 \eqno(28.a)$$
$$R_i(\{ s_{j+1} \})/R_i (\{ s_j \} )_{min} = 1 + \eta_2 \eqno(28.b)$$

\noindent $\eta_1$ and $\eta_2$ being positive and non-infinitesimal. As $j 
\to
\infty$, the interval of variation of $R_i (\{ s_j \} )$ we tend to zero, 
and a
value $R_i^{\infty}$ independent of $\{ s_j \}$ will be obtained for $R_i \{ 
s_j
\}$. \par

However, there may be a snag if the weight of $R_i(\{ s_j \})_{max}$ or $R_i(
\{ s_j
\})_{min}$ is infinite with respect to the sum of the weights of the other 
values of
$R_i(\{ s_j \})$ {\it for some} $\{ s_{j+1} \}$, because (28) may not hold. 
In the
next section, we study the weight-ratio 
$$W_{m_2}^{m'}(\{ s_j \}_2, \{ s_{j+1} \})/W_{m_1}^{m'}(\{ s_j \}_1, \{ 
s_{j+1}
\})$$

\noindent of weights corresponding to $R_i^{m_2}(\{ s_j \}_2)$ and 
$R_i^{m_1}(\{ s_j
\}_1)$ in the evaluation of $R_i^{m'}(\{ s_{j+1} \})$ and prove that when a 
certain
constraint (easy to implement) on the construction of the $V_j$'s is 
imposed, it is
independent of $\{ s_{j+1} \}$. This allows us to conclude that either $
\eta_j$ is
finitely different from zero or that when it is infinitesimal, it is so for 
any $\{
s_{j+1} \}$, ensuring convergence.

 \medskip \noindent {\bf B - {\it Self-avoiding paths}}\par

\noindent {\bf Definition} \par

A path $P(v_1 , v_n)$ is defined as the succession of propagators $(v_1 
v_2)$, $(v_2
v_3)$, $\cdots$ , $(v_{n-1} v_n)$ linking $v_1$ to $v_n$, \ $v_1$, $v_2$, $
\cdots$ ,
$v_n$ being $n$ vertices on $G$. In a self-avoiding path $v_1$, $v_2$, $
\cdots$ ,
$v_n$ are all different vertices. \par 

A closed path is constructed when $v_1$ and $v_n$ are the same vertex. A 
loop is a
self-avoiding closed path. $\sq$ \par

The main tool we will be using now is the fact that, taking a vertex $v_k$ 
at one end
of the propagator $k$ and a vertex $v_i$ at one end of the propagator $i$, 
for each
spanning tree ${\cal T}$ on $G$, there is one path, and only one, on ${\cal 
T}$,
binding $v_k$ and $v_i$. Furthermore, {\it this path is self-avoiding}, it 
goes
through each vertex it is incident with only once. So the idea is to 
construct all
spanning trees on $G$ by beginning to construct all self-avoiding paths 
$P(v_i,
v_k)$ on $G$ binding $v_i$ and $v_k$. Two spanning trees of $G$ having a
different path $P(v_i, v_k)$ are necessarily different. (If that were not 
the case,
we would have two different paths $P(v_i, v_k)$ on the same tree, giving a 
loop on
this tree which is forbidden). Of course, for every such path there exist 
many
spanning trees obtained by sprouting branches out of the path. In fact, 
counting the
number of trees associated with one path is the same as counting the number 
of ways
branches can be sprouted out of this path. However, in a first step, we will
concentrate our attention on the paths $P(v_i, v_k)$ themselves, taking into 
account
the effect of branches later on, i.e. in the next section. \par 

Now, let us consider the sum over all paths, each one being weighted by the 
product of
all $\bar \alpha_l^{-1}$ belonging to the propagators along it. We denote by 
$P(i,
s)$ a path $P(v_i, s)$ which {\it goes through the propagator} $i$ and by $P(
\bar i,
s)$ a path which {\it does not go through $i$}, both paths relating $v_i$ 
and $s$.
Then, the sums over all $P(i, v_k)$ and $P(\bar i, v_k)$ can be written (the 
sum over
paths $P$, $\displaystyle{\sum_P}$, is a multiple sum, a summation being 
made for
each path $P(s_p, s_{p+1})$),  

$$\sum_P P(i, v_k) = \sum_P
\sum_{s_i,...,{s_{l}}_{k-1}} P(i, s_1) P(s_1, s_2) ... P({s_{l}}_{k-1}, v_k) 
\eqno
(29.a)$$  

$$\sum_P P(\bar i, v_k) = \sum_P \sum_{s_i,...,{s_l}_{k-1}} P(\bar i,
s_1) P(s_1, s_2) ... P({s_l}_{k-1}, v_k) \eqno (29.b)$$

\noindent $s_1, s_2, ..., {s_l}_{k-1}$ belonging to the ensemble of the
border-vertices of closed volumes\break \noindent $V_1, V_2, ..., 
{V_l}_{k-1}$ such
that  $$V_1 \subset V_2 \subset ... \subset {V_l}_{k-1} \eqno(30)$$

\noindent It is important to note that a path $P(s_j, s_{j+1})$ will
{\it go out of} $V_j$ {\it at} $s_j$ {\it for} {\it the first
time} but will otherwise be entirely contained in $V_{j+1}$. Moreover,
$P(s_j, s_{j+1})$ can re-enter $V_j$ and go out again at some vertex $s'_j$.
We also will take $V_1$ to have a border at a finite distance of $v_i$, and 
the border
of $V_{j+1}$ to be at a finite distance of $V_j$ {\it in terms of the 
minimum number
of propagators separating them}.\par

Let us now define this notion of distance on a graph. We are going to embed 
$G$ in an
Euclidean space where each propagator has a definite length (this is the 
purpose of
this embedding). Therefore, any path will have a definite length in this 
Euclidean
space. The length of the path with the least number of propagators will be 
the distance
between two vertices on $G$. \par

Let us also remark that, according to our above description, the distance 
between two
vertices measured in Euclidean space and measured in the least number of 
propagators
joining them have a priori {\it no monotonicity relation
between them}, because the lengths of propagators can vary from one part to 
another
part of $G$. We now define in a constructive way the volumes $V_j$. \par

\medskip \noindent {\bf C - {\it Construction of the $V_j$'s}}\par
  
We suppose that $G$ is embedded in a 3-dimensional $R_3$ space. Then, we 
define the
sphere $S_1$, i.e. the ensemble of points {\it within} a certain radius, 
centered at
the vertex $v_i$, provided a metric has been defined in $R_3$. The radius of 
$S_1$
will be taken such that $S_1$ contains a finite number of propagators. In
general, the two-dimensional border-surface of $S_1$ cut through propagators 
of $G$.
We then deform continuously $S_1$ inwards in such a way that its border 
slides along
the cut propagators until vertices attached to these propagators are met. 
\par

\noindent {\bf Definition} \par

The border of $V_1$ is the deformed border of $S_1$. $\sq$ \par

As we suppose that $v_i$ is inside $S_1$, it will also be inside $V_1$. Note 
that
the border of $V_1$ may also contain entire propagators linking two vertices 
on its
boundary. All the vertices and propagators belonging to the deformed sphere 
$S_1$ will
belong to $V_1$. \par

Constructing $V_2$, we start with a sphere $S_2$ of radius larger than that 
of $S_1$. Then,
we deform $S_2$ along the propagators cut by its border-surface until we 
meet the vertices
attached to these propagators. Now, we are tempted to define $V_2$ in the 
same way as we did
for $V_1$, by taking its border to coincide with the deformed sphere $S_2$. 
However, we will
do so only in the case where the part of $G$ between the borders of the 
deformed $S_2$ and
$S_1$ is connected. In general, we expect that, in fact, {\it there will be 
several connected
pieces of $G$ between the borders of $S_2$ and $S_1$}. For each of these 
pieces we
will have a corresponding connected piece of $V_2 - V_1$ (meaning $V_2$ from 
which
$V_1$ has been subtracted) containing all the vertices and propagators of 
this
connected piece of $G$. Let us consider in more details a given connected 
piece in
$V_2 - V_1$. \par

Such a piece, let us call it $(V_2 - V_1)_C$ has a boundary formed by three 
pieces
\par 

\hskip 1 truecm i) a piece on the border of the deformed sphere $S_2$, which
itself has a closed curve $C_2$ as boundary, \par

\hskip 1 truecm ii) a piece on the border of the deformed sphere $S_1$, which
itself has a closed curve $C_1$ as boundary, \par  

\hskip 1 truecm iii) a cylindrical piece having $C_1$ and $C_2$ as 
boundaries. It
should not be crossed by $G$. \par

In short, each connected $(V_2 - V_1)_C$ will enclose a corresponding 
connected
part of $G$ in $V_2 - V_1$, let us call it $G_C$. Then, $C_2$ encloses all 
the
vertices and propagators of $G_C$ on the surface of the deformed $S_2$. In 
the
same manner $C_1$ encloses all the vertices and propagators of $G_C$ on the
border of $V_1$ (which according to the last definition is the surface of the
deformed $S_1$). Note that there may be, in some cases, no vertex of $G_C$ 
either
in the surface enclosed by $C_1$ or in the surface enclosed by $C_2$, due to 
the
topology of $G$ ($G$ may not go through the surface of the deformed $S_2$ at 
the
border of $(V_2 - V_1)_C$, or through the piece of border of $V_1$ in common 
with
that of $(V_2 - V_1)_C)$. \par

Another condition we have on the boundary of $(V_2 - V_1)_C$ is that it 
should
not cross the boundary of another connected piece in $V_2 - V_1$. In that 
way,
two different connected volumes of $V_2 - V_1$ do not overlap. \par 

{\it This condition and the properties i), ii) and iii) above define the boun
\-da\-ry
of one connected part $(V_2 - V_1)_C$}. \par 

For the following volumes $V_3, \cdots , V_j$, the constructive process we 
just
described for $V_2$ repeats itself, each connected piece of $V_j - V_{j-1}$ 
being
defined through its boundaries. It may happen that two successive $V_{j-1}$, 
$V_j$
have some coinciding part of their borders (on a common part of the deformed
$S_{j-1}$ and $S_j$) because the density of propagators may be much larger 
in other
parts of $V_j - V_{j-1}$. This ends our construction of the volumes $V_j$. 
\par 

\medskip \noindent {\bf D - {\it An averaging theorem}}\par

  We then have an unequivocal definition for the paths $P(v_i, v_k)$. In ge
\-ne\-ral
the ratio $P(i, s_1)/P(\bar i, s_1)$ depends on $s_1$. However, considering 
sums over
paths, we want to prove that  
$$R_i(s_j) = \sum_P P(i, s_j)/\sum_P P(\bar i, s_j) \eqno(31)$$ 

\noindent {\it tends to a value independent of} $s_j$ {\it as} $j \to
\infty$. This is the clue for deriving (24) and thereby the factorization
property. (It will also be proven in section 5 that when $k$ is outside 
$V_j$, $j \to
\infty$, for a given $k$, the factors
$s_{C_k} \nu^{-1} ({\cal C}_k)$ and $s_{{\cal C}_k^i} \nu^{-1} ({\cal 
C}_k^i)$ are
equal). Now we want to prove the following lemma \par

\noindent {\bf Lemma 1} \par 

If $P(s_j, s_{j+1})$ never returns in $V_j$ then,
$$R_i(s_j)_{min} \leq R_i(s_{j+1}) \leq R_i(s_j)_{max} \eqno(32)$$

\noindent where $R_i(s_j)_{min}$ and $R_i(s_j)_{max}$ are respectively the
minimum and the maximum values of $R_i(s_j)$ for those $s_j$ coupled to
$s_{j+1}$ by at least one path $P(s_j, s_{j+1})$. \par

\noindent {\bf Proof} \par
 
As a first remark, we can factorize the expression for
$\displaystyle{\sum_P P(i, s_{j+1})}$,

$$\sum_{s_j} \sum_P P(i, s_j) P(s_j, s_{j+1}) = \sum_{s_j}
\lbrack \sum_P P(i, s_j) \rbrack \ \lbrack \sum_P P(s_j, s_{j+1}) \rbrack$$

\noindent because the paths $P(s_j, s_{j+1})$ being entirely in $V_{j+1} -
V_j$ never interact with the paths $P(i, s_j)$ contained in $V_j$. Therefore,
using (31) we can write  
$$\vbox{\eqalignno{
R_i(s_{j+1}) &= \lbrace \sum_{s_j} \lbrack
\sum_P P(i, s_j) \rbrack \ \lbrack \sum_P P(s_j, s_{j+1}) \rbrack \rbrace
/ \cr 
&\lbrace \sum_{s_j} \ \lbrack \ \sum_P P(\bar i, s_j) \rbrack \
\lbrack \ \sum_P P(s_j, s_{j+1}) \rbrack \rbrace \cr
&= \lbrace \sum_{s_j}
R_i(s_j) \ \lbrack \sum_P P(\bar i, s_j) \rbrack \lbrack \sum_P P(s_j,
s_{j+1}) \rbrack \rbrace / \cr 
&\lbrace \sum_{s_j} \ \lbrack \sum_P P(\bar i, s_j)
\rbrack \lbrack \sum_P P(s_j, s_{j+1}) \rbrack \rbrace &(33) \cr
}}$$

\noindent We see that $R_i(s_{j+1})$ is an average of $R_i(s_j)$ for those
$s_j$ coupled to $s_{j+1}$ and therefore (32) is true. $\sq$ \par 

Our goal is, of course, a repeated use of (32) and as $j$ grows we expect 
$R_i(s_j)$
to become independent of $s_j$. However, we have restricted the paths $P(s_j,
s_{j+1})$ to be in $V_{j+1} - V_j$ in order to avoid an interaction with the
paths $P(v_i, s_j)$. If we allow such an interaction to occur, i.e. if 
$P(s_j,
s_{j+1})$ returns in $V_j$, we have to distinguish between the different
topologies of $P(s_j, s_{j+1})$. That is, we have to cut $P(s_j, s_{j+1})$ in
parts which stay in $V_j$ and parts which stay in $V_{j+1} - V_j$
{\it in order to be able to factorize the sum over paths}. Let us
call $s_j^0$ the first vertex of $P(s_j, s_{j+1})$ and $s_j^1, s_j^3, ...,
s_j^{2m-3}$ the vertices where $P(s_j, s_{j+1})$ re-enters $V_j$. (At 
$s_j^2$,
$s_j^4$, ..., $s_j^{2m-2}$ $P(s_j, s_{j+1})$ goes out of $V_j$). We denote by
$\lbrace s_j^l \rbrace$ the ensemble $s_j^0, s_j^1, s_j^2, ..., s_j^{2m-2}$ 
and
$M_j$ the maximum number of connected parts of $P(i, s_{j+1})$ in $V_j$. 
Then,
$\displaystyle{\sum_P} \ P(i, s_{j+1})$  can be written

$$\sum_P P(i, s_{j+1}) = \sum_{s_j^0} \ \lbrack \sum_P P(i, s_j^0) \rbrack \ 
\lbrack
\sum_P P(s_j^0, s_{j+1}) \rbrack$$
 $$ + \sum_{m=2}^{M_j} \ \sum_{\lbrace s_j^l \rbrace} \sum_{\lbrace s_j^{2l},
s_j^{2l-1} \rbrace}$$  
$$\lbrack \sum_P P(i, s_j^0) \prod_{l=1}^{m-1} P(s_j^{2l-1}, s_j^{2l}) 
\rbrack$$
$$\lbrack \sum_P P(s_j^{2m-2}, s_{j+1}) \prod_{l=1}^{m-1} P(s_j^{2l-2},
s_j^{2l-1}) \rbrack \eqno(34)$$

\noindent where the paths $P(i, s_j^0)$, $P(s_j^{2l-1}, s_j^{2l})$ are
contained in $V_j$ and the paths $P(s_j^0, s_{j+1})$, $P(s_j^{2l-2},
s_j^{2l-1})$ and $P(s_j^{2m-2}, s_{j+1})$ have all their propagators 
contained
in $V_{j+1} - V_j$. $\{s_j^{2 \ell}, s_j^{2 \ell - 1} \}$ is the ensemble of
couples $(s_j^{2 \ell}, s_j^{2 \ell - 1})$ on $P(s_j, s_{j+1})$. The first 
term in
(34) stands for the case where $P(s_j, s_{j+1})$ never returns in $V_j$ and 
therefore
corresponds to the case of Lemma 1 ($m$ = 1). The following terms describe 
the case
where $P(s_j, s_{j+1})$ returns in $V_j$ ($m \geq 2$). Then, for a given $m$,
$\lbrace s_j^l \rbrace$, $\lbrace s_j^{2l}, s_j^{2l-1} \rbrace$ the 
summation over
paths has been factorized into paths contained in $V_j$ and paths contained 
in
$V_{j+1} - V_j$. Notice that the couples $(s_j^{2l}, s_j^{2l-1})$, in fact, 
select an
order among the $s_j^l$'s, their order along the path $P(s_j^0, s_{j+1})$. 
However,
the sum over all orders can be factorized as a sum over all orders of paths 
in $V_j$
times a sum over all orders of paths in $V_{j+1} - V_j$. So, {\it we could 
dispense
ourselves of tracking down the order of the $s_j^l$'s through the couples 
$(s_j^{2l},
s_j^{2l-1})$. We do so only in order to make clear how the summation over 
paths is
done}. Later on, {\it in the case of spanning trees this order will be 
meaningless}
and it will not appear in the summation over all trees. Let us now define by 
(for $m
\geq$ 2)  
$$R_i^m(s_j) = \lbrack \sum_P P(i, s_j^0) \prod_{l=1}^{m-1}
P(s_j^{2l-1}, s_j^{2l} )\rbrack /$$ $$\lbrack \sum_P P(\bar i, s_j^0)
\prod_{l=1}^{m-1} \bar P(s_j^{2l-1}, s_j^{2l} ) \rbrack \eqno(35)$$

\noindent the ratio of the part of $\displaystyle {\sum_P} P(i, s_{j+1})$ in
$V_j$ over the part of $\displaystyle {\sum_P} P(\bar i, s_{j+1})$ \ which
also is in $V_j$ ; $m$, $\lbrace s_j^l \rbrace$ and the couples $\lbrace 
s_j^{2l}, s_j^{2l-1}
\rbrace$ are given.

\par \noindent $\bar P$ helps to distinguish the paths which are
associated with $P(\bar i, s_j^0)$. Of course, $R_i(s_j) \equiv
R_i^1(s_j)$. \par

We remark that the sums over $P$ in (35) have {\it not} been factorized. This
is, of course, because the paths in $V_j$ have to avoid each other, so that a
configuration of one of them affect the summation over the others. Then, we 
have the
following theorem \par 
 
\noindent {\bf Theorem 1} 
$$R_i^m(s_j)_{min} \leq R_i^{m'}(s_{j+1}) \leq R_i^m(s_j)_{max} \eqno(36)$$

\noindent where $R_i^m(s_j)_{min}$ and $R_i^m(s_j)_{max}$ are respectively 
the
minimum and the maximum values of $R_i^m(s_j)$ viewed as a function of $m$,
$\lbrace s_j^l \rbrace$ and $\lbrace s_j^{2l}, s_j^{2l-1} \rbrace$, and for 
those
$s_j^l$ on at least one path $P(i, s_{j+1})$. \par
\vfill \supereject 
\noindent {\bf Proof} \par

For the sake of simplicity of notation {\it we will make
implicit the sum} $\displaystyle {\sum_P}$ {\it each time the symbol $P$
appears}. Then define
$$P^m(i, s_j) = P(i, s_j^0) \prod_{l=1}^{m-1} P(s_j^{2l-1},
s_j^{2l}) \eqno(37.a)$$

$$P^m(\bar{i}, s_j) = P(\bar{i}, s_j^0) \prod_{l=1}^{m-1} \bar{P}(s_j^{2l-1},
s_j^{2l}) \eqno(37.b)$$

\noindent where the symbol $\bar{P}$ is for a path which is associated with
$P(\bar{i}, s_j^0)$ and could differ from a path $P$ because they are 
returning
in $V_1$ where the paths $P(i, s_j^0)$ and $\bar{P}(i, s_j^0)$ are 
different. Writing
$P^{m'}(i, s_{j+1})$ we obtain (we remark that, except for $P(i, s_j^0)$ and
$\displaystyle{\prod_{l=1}^{m-1}} P(s_j^{2l-1}, s_j^{2l})$ which are 
confined in
$V_j$, all other $P$'s are in $V_{j+1} - V_j$) (see fig. 1)
$$\vbox{\eqalignno{
P^{m'}(i, s_{j+1}) = &\sum_{s_j^0} P(i, s_j^0) P(s_j^0, s_{j+1}^0)
\prod_{l=1}^{m'-1} P(s_{j+1}^{2l-1}, s_{j+1}^{2l}) \cr 
&+ \sum_{m=2}^{M_j} \sum_{\lbrace s_j^l \rbrace} P(i,s_j^0) \prod_{l=1}^{m-1}
P(s_j^{2l-1}, s_j^{2l})\ . \cr 
&\sum_{\lbrace l_t \rbrace} \prod_{t=1}^{t_j} P(s_j^{2l_t-2}, s_j^{2l_t-1})
\ . \cr
&\sum_{\lbrace l_u \rbrace} \prod_{u=1}^{u_j} P(s_j^{l_u},s_{j+1}^{l_u})\ . 
\cr
&\sum_{\lbrace l_v \rbrace} \prod_{v=1}^{v_{j+1}} P(s_{j+1}^{2l_v-2},
s_{j+1}^{2l_v-1}) &(38) \cr
}}$$
 
\noindent with 
$$m - t_j + v_{j+1} = m' \eqno(39.a)$$
 
$$\lbrace s_j^l \rbrace = \lbrace s_j^{2l_t-2} \rbrace \cup \lbrace 
s_j^{2l_t-1}
\rbrace \cup \lbrace s_j^{l_u} \rbrace \eqno(39.b)$$

The first term in (38) stands for the case where the connected
component starting from $v_i$ of $P^{m'}(i, s_{j+1})$, $P(i, s_j^0) \ 
P(s_j^0, s_{j+1}^0)$, goes out of $V_j$ at $s_j^0$ and stays in
$V_{j+1} - V_j$, with all other connected components $P(s_{j+1}^{2l-1},
s_{j+1}^{2l})$ also staying in $V_{j+1} - V_j$. In this case, there is
only one connected part in $V_j$, $P(i, s_j^0)$, of $P^{m'}(i,
s_{j+1})$ and therefore this corresponds to $m$ = 1. The following
terms in (38) describe the other cases with $m$ components $(m \geq 2)$
in $V_j : P(i, s_j^0)$ with $m$ - 1 other components $P(s_j^{2l-1},
s_j^{2l})$. The connected paths $P(s_j^{2l_t-2}, s_j^{2l_t-1})$ are
never incident with the border of $V_{j+1}$ and end up at the border of
$V_j$. The connected paths $P(s_j^{lu}, s_{j+1}^{lu})$ start on the
border of $V_j$ and end up on the border of $V_{j+1}$ while the
connected paths $P(s_{j+1}^{2l_v-2}, s_{j+1}^{2l_v-1})$ start and end up
on the border of $V_{j+1}$ with no incidence with the border of $V_j$.
\par 

We remark that for $m'$ = 1 we recover expression (34) because
the products over the paths $P(s_{j+1}^{2l-1}, s_{j+1}^{2l})$ and
$P(s_{j+1}^{2l_v-2}, s_{j+1}^{2l_v-1})$ do not exist in that case and
the product over $P(s_j^{lu}, s_{j+1}^{lu})$ is replaced by one unique
term $P(s_j^{2m-2}, s_{j+1})$ in (34). \par 

The relation (39.a) is obtained by counting the number of vertices of 
$P^m(i, s_j)$ on
the border of $V_j$ which is equal to 2$m$ - 1 and the number of vertices
of $P^{m'}(i, s_{j+1})$ on the border of $V_{j+1}$ which is equal to
2$m'$ - 1 and calculating the difference 2($m'$ - $m$). This difference
comes from the connected paths $P(s_{j+1}^{2l_v-2}, s_{j+1}^{2l_v-1})$
which contributes $2 v_{j+1}$ and $P(s_j^{2l_t-2}, s_j^{2l_t-1})$ which
contributes $- 2 t_j$ to it. The same expression as (38) is obtained for
$P^{m'}(\bar i, s_{j+1})$ by replacing $P(i, s_j^0)$ by $P(\bar i, s_j^0)$ 
and
$P(s_j^{2l-1}, s_j^{2l})$ by $\bar P(s_j^{2l-1}, s_j^{2l})$. Then, writing 
(sums
over $P$, again, are implicit) $$R_i^{m'}(s_{j+1}) = P^{m'}(i, 
s_{j+1})/P^{m'}(\bar i,
s_{j+1}) \eqno(40)$$

\noindent and using (35) in it we obtain $R_i^{m'}(s_{j+1})$ as an average of
$R_i^m(s_j)$ function of the $s_j$'s and $m$, the average being the result
of a sum over $m$, $\lbrace s_j^l \rbrace$ and the sets of couples $\lbrace
s_j^{2l-1}, s_j^{2l} \rbrace$, $\lbrace s_j^{2l_t-2}, s_j^{2l_t-1} \rbrace$,
$\lbrace s_j^{lu} , s_{j+1}^{lu} \rbrace$ and $\lbrace s_{j+1}^{2l_v-2} ,
s_{j+1}^{2l_v-1} \rbrace$ aside from the sum over paths $P$. However, once
the set $\lbrace s_j^l \rbrace$ is given the partition of the paths in
$V_{j+1} - V_j$ does not depend on that of the paths in $V_j$ because there
is no interaction  between them. Therefore the sum over all couples can be
factorized out as it was already the case for the sum over the couples
$\lbrace s_j^{2l-1}, s_j^{2l} \rbrace$. Consequently, the only variables
which are necessary to retain are $m$ and the set $\lbrace s_j^l \rbrace$ in
the functional dependence of $R_i^m(s_j)$. As said earlier we retain the
dependence on the couples $\lbrace s_j^{2l}, s_j^{2l-1} \rbrace$ in order to
remind ourselves that we had to sum over all orders, this sum being, again,
factorizable. Therefore the relation (36) follows. $\sq$ \par \smallskip

Again, a repeated use of (36) will allow us to make all the ratios 
$R_i^m(s_j)$
converge towards one value independent of $s_j$ as $j \to \infty$. \par 

However, we have to be careful about two problems potentially hampering the 
efficiency
of this uniformization of $R_i^m(s_j)$. \par 

a) As long as $R_i^m(s_j)$ has not converged, both inequalities (the same as 
(28.a) and
(28.b))  $$R_i^{m'}(s_{j+1})/R_i^m(s_j)_{max} < 1 - \eta_1
\eqno(41.a)$$
$$R_i^{m'}(s_{j+1})/R_i^m(s_j)_{min} > 1 + \eta_2 \eqno(41.b)$$

\noindent should be satisfied, $\eta_1$ and $\eta_2$ being two positive
non-infinitesimal constants. \par 

b) $G$ could have the topology of a tree-like structure or "polymer", i.e.
many branches could stem out of $V_1$ and $V_{j+1} - V_j$ would be multiply
connected. Remember that in Lemma 1 and Theorem 1, we have the restriction 
that
$s_j$'s should be on at least one path $P(i, s_{j+1})$. Then, a path going in
some connected part of $V_{j+1} - V_j$ could never go in another connected 
part
of $V_{j+1} - V_j$ because in order to do so, it would have to return in
$V_1$ which may be impossible because it would have to go through vertices
with which it is already incident. Then, due to this finite volume effect,
most paths going along one branch of $V_{j+1} - V_j$ would never go in
another branch. This leaves the possibility for $R_i^m(s_j)$ of evolving 
towards
a different value along each branch of this tree-like structure, instead
of a unique value as we wanted to show. \par 

However, for spanning trees on $G$ this difficulty is easily removed. The 
reason for
this is clear : a spanning tree of $G$ is incident with every vertex of $G$. 
Then, we
can think that a spanning tree on $G$ is composed of one path $P(i, 
s_{j+1})$ and a
myriad of paths stemming out of it, the ensemble of all paths going through
all vertices of $V_{j+1}$, and then through all connected domains of $V_{j+1}
- V_j$. For trees, a separate evolution of the quantity corresponding to 
$R_i^m(s_{j+1})$ along branches of $V_{j+1} - V_j$ is therefore prohibited.
That is the essential difference between paths and spanning trees in our
approach.\par 

In the next section we also show how the problem a) is solved by insuring 
the validity
of (41.a) and (41.b).\par \medskip

\noindent {\bf 4. Trees on G} \medskip 

\noindent {\bf A - {\it Convergence of the iteration of mean-value 
operation}}\par 

Let us consider $R_i(s_{j+1})$ as given by the mean-value expression (33). 
For
each $s_j$, $R_i(s_j)$ is multiplied by a weight $W(s_j, s_{j+1}) =
\displaystyle{\sum_P} P(\bar{i}, s_j) \ P(s_j, s_{j+1})/\displaystyle{
\sum_P} P(\bar{i},
s_{j+1})$. Either (41.a) or (41.b) may be violated if and only if in the sum
$\displaystyle{\sum_{s_j}} R_i(s_j) W(s_j, s_{j+1})$ the weight $W(s_j, 
s_{j+1})$ associated
with $R_i(s_j)_{max}$ or $R_i(s_j)_{min}$ is infinite with respect to the 
sum of weights of
any other value for $R_i(s_j)$. (By infinite, we mean that the ratios should 
be infinite).
\par Indeed, let us verify this and use (33), isolating the contribution of
$R_i(s_j)_{max}$, $s_{jm}$ being the vertex for which $R_i(s_{jm}) =
R_i(s_j)_{max}$

$$R_i(s_{j+1}) = [ R_i(s_j)_{max} \ W(s_{jm} , s_{j+1}) + \sum_{s_j \not=
s_{jm}} R_i(s_j) \ W(s_j , s_{j+1}) ]/$$ $$[ W (s_{jm} , s_{j+1}) +
\sum_{s_j \not= s_{jm}} W(s_j , s_{j+1}) ] \ \ \ . \eqno(42)$$

We will take $R_i(s_j)/R_i(s_j)_{max}$ for $s_j \not= s_{jm}$ to have a {\it 
finite}
(non-infinitesimal) difference with one ({\it otherwise convergence
to a unique value is already achieved}) and simplify the notation, defining 
$$W_m = W \left ( s_{jm} , s_{j+1} \right ) \eqno(43.a)$$

$$W = \sum_{s_j \not= s_{jm}} W(s_j , s_{j+1}) \ \ \ .  \eqno(43.b)$$

\noindent Then,
$$R_i(s_{j+1})/R_i(s_j)_{max} = \left [ W_m \ R_i(s_j)_{max} + W 
\bar{R}_i(s_j)
\right ]/$$ $$\left [ (W_m + W) R_i (s_j)_{max} \right ] \eqno(44)$$

\noindent with
$$W \bar{R}_i(s_j) = \sum_{s_j \not= s_{jm}} R_i(s_j) \ W(s_j, s_{j+1}) 
\eqno(45)$$

\noindent $\bar{R}_i(s_j)$ being the mean-value of $R_i(s_j)$ for $s_j \not=
s_{jm}$. We have
$$\bar{R}_i(s_j) = (1 - \eta) \ R_i(s_j)_{max} \eqno(46)$$

\noindent $\eta > 0$, and {\it non-infinitesimal}, following our assumption 
that
$R_i(s_j)/R_i(s_j)_{max}$ is finitely different from one for $s_j \not=
s_{jm}$. Hence, (44) gives 
$$R_i(s_{j+1})/R_i(s_j)_{max} = 1 - \eta/(1 + W_m/W)
\eqno(47)$$

\noindent {\it which shows that} (41.a) {\it is satisfied provided the ratio
$W_m/W$ is not infinite}. Of course, the same sort of reasoning is also valid
for showing that (41.b) is violated only if the weight of $R_i(s_j)_{min}$ is
infinite with respect to the sum of the other weights. \par
  
Of course, it could also be that while having a finite weight, either 
$R_i(s_j)_{max}$
or $R_i(s_j)_{min}$ dominates the sum because the number of $s_j$'s with 
$R_i(s_j) \not=$ (either $R_i(s_j)_{max}$ or $R_i(s_j)_{min}$) is 
infinitesimal
with respect to the number of $s_j$'s with $R_i(s_j)$ = (either 
$R_i(s_j)_{max}$
or $R_i(s_j)_{min}$). However, in this particular case, the range of 
variation of
$R_i(s_{j+1})$ would be infinitesimal and its convergence to value
independent of $s_{j+1}$ insured, which is what we want to demonstrate to
be valid in any case. \par 

Then, to obviate the difficulty mentioned above in the case of infinite 
weight-ratio
$W_m/W$ for either $R_i(s_j)_{max}$ or $R_i(s_j)_{min}$ we may show that 
taking
$s_{1j}$ and $s_{2j}$ separated by a finite number of propagators in 
$V_{j+1} - V_j$
as well as in $V_j$ the weight associated with $R_i(s_{2j})$ can be obtained 
from
the weight associated with $R_i(s_{1j})$ by multiplying it by a finite 
factor. If
$R_i(s_{1j})$ is $R_i(s_j)_{max}$ or $R_i(s_j)_{min}$ and 
$R_i(s_{2j})/R_i(s_{1j})$
is finitely different from one, the ratio $W_m/W$ will then stay finite and 
the
relations (41) will be valid. This can be easily understood because the
corresponding paths $P(\bar{i}, s_{j+1})$ can be obtained from each other by 
a
``local'' deformation, i.e. by substituting only a finite number of 
propagators.
\par

In the following, we will consider a proof of the validity of (41) inspired 
from
this idea, but specific to trees. Therefore, we will need to translate the
language adopted for paths and multi-paths into the one adopted for spanning
trees and multiple-spanning trees. When we consider a spanning tree on $G$, 
the
part of $G$ contained in $V_j$ will in general be a spanning $m$-tree in 
$V_j$,
i.e. a spanning tree in $V_j$ from which $m$ propagators have been removed. 
The
vertices of the border of $V_j$ will be separated into $m$ sub-sets $
\{s_j^{m_c}\}$
(because each sub-tree has to be incident with the border of $V_j$ on the
border of the deformed $S_j$ in order to be connected to the rest of the
spanning tree on $G$), each sub-set $\{s_j^{m_c}\}$ belonging to one of the 
$m$
sub-trees belonging to the spanning $m$-tree in $V_j$. There will also be
sub-trees in $V_{j+1} - V_j$. Some (or all) of them will connect to a 
sub-tree
in $V_j$ to form, as a whole, a spanning $m'$-tree in $V_{j+1}$, i.e. a
$m'$-tree which is incident with all the vertices contained in $V_{j+1}$. In
the same way the vertices on the border of $V_{j+1}$ will be divided into 
$m'$
sub-sets, each belonging to a sub-tree of a spanning $m'$-tree in $V_{j+1}$.
Therefore, in analogy with the multi-paths situation, we will write
$$R_i^m(\{s_j\}) = \sum_{T^m} T^m(i, \{s_j\})/\sum_{T^m} T^m(\bar{i}, \{ s_j
\}) \eqno(48)$$

\noindent where $T^m(i, \{s_j\})$ is the weight for a spanning $m$-tree
in $V_j$ going through the propagator $i$ and $T^m(\bar{i}, \{s_j\})$ is the
weight of a spanning $m$-tree in $V_j$ not going through $i$. \par

As for multi-paths the only information conveyed from the structure of
$m$-trees in $V_j$ to the sub-trees in $V_{j+1} - V_j$ is the partition
$\{s_j\}$ of the vertices on the part of the border of $V_j$ which is on the
border of the deformed sphere $S_j$. This partition is common to the 
$m$-trees
of weight $T^m(i, \{s_j\})$ and weight $T^m(\bar{i}, \{s_j\})$ and is the
ensemble of all sub-sets $\{s_j^{m_c}\}$. We write in analogy with (33)
$$R_i^{m'}(\{s_{j+1}\}) = \sum_{\{s_j\}} R_i^m(\{s_j\}) \ W_m^{m'}(\{s_j\},
\{s_{j+1}\}) \eqno(49.a)$$

$$W_m^{m'}(\{s_j\} , \{s_{j+1}\}) = \sum_{T^m} T^m (\bar{i}, \{s_j\})
\sum_{T^n} T^n(\{s'_j\} , \{ s_{j+1}\})/N_{m'} \eqno(49.b)$$

$$N_{m'} = \sum_{T^{m'}} T^{m'} (\bar{i}, \{s_{j+1}\}) \eqno(49.c)$$

\noindent where $T^n(\{s'_j \} , \{s_{j+1}\})$ is the weight of a spanning
$n$-tree in $V_{j+1} - V_j$ {\it which together with a $m$-tree in $V_j$ 
forms
a spanning $m'$-tree in $V_{j+1}$}. $\{s'_j\}$ may not be identical to
$\{s_j\}$ because not all vertices in $\{s_j\}$ may be incident with a
$n$-tree in $V_{j+1} - V_j$. $N_{m'}$ is the sum over the weights of all 
spanning $m'$-trees
in $V_{j+1}$, $T^{m'}(\bar{i}, \{s_{j+1}\})$, which do not go through $i$ and
correspond to $\{s_{j+1}\}$. \par

Some partitions $\{s_j\}_M$ correspond to an extremum value
$R_i^m(\{s_j\}_M)$ and to ensure the validity of (41) we have to ensure
that the weight $W_{m_M}^{m'}(\{s_j\}_M , \{s_{j+1}\})$ corresponding
to these partitions does not become infinite relative to the sum of weights 
of
the other partitions. Or, alternatively if that case arises, showing that the
ratio
$$\sum_{\{s_j\}_M} W_{m_M}^{m'}(\{ s_j \}_M , \{s_{j+1} \}/
\sum_{\{s_j\}} W_m^{m'} (\{s_j \}, \{ s_{j+1}\}) \eqno(50)$$  

\noindent stays infinite whatever $\{s_{j+1}\}$ is also solves our problem. 
Then, we
would already have convergence to a unique value $R_i^{m'}$ equal to
$R_i^m(\{s_j\}_M)$. Our strategy will consist in building a constructive
procedure which allows us to calculate the ratio of two weights
$$W_{m_2}^{m'}(\{s_j\}_2 , \{s_{j+1}\})/W_{m_1}^{m'}(\{ s_j \}_1, \{s_{j+1}
\})
\eqno(51)$$

\noindent for two different partitions $\{s_j\}_1$ and $\{s_j\}_2$, the 
partition
$\{s_{j+1}\}$ remaining the same. $\{s_{j+1}\}$ is the partition of vertices 
on the
part of the border of $V_{j+1}$ on the deformed sphere $S_{j+1}$. It 
consists in
sub-sets $\{s_{j+1}^{m'_c}\}$, $(m'_c = 1 , \cdots , m')$ of vertices on a 
sub-tree
of a spanning $m'$-tree in $V_{j+1}$. Now, $\{s_j\}_2$ can always be 
obtained from
$\{s_j\}_1$ by {\it a series of minimal modifications}, each of which 
consists in
cutting in $V_{j+1} - V_j$ a self-avoiding path connecting two vertices 
$s_{1j}$
and $s_{2j}$ belonging to those forming $\{s_j\}$ instead of cutting a path
staying $V_j$ relating the same vertices $s_{1j}$ and $s_{2j}$. Thus, in a
minimal modification a propagator in $V_{j+1} - V_j$ on a $m'$-tree in 
$V_{j+1}$
is replaced by a propagator in $V_j$, obtaining another $m'$-tree in 
$V_{j+1}$
but with the same $\{s_{j+1}\}$ because $s_{1j}$ and $s_{2j}$ (and the 
vertices
$s_{j+1}$ connected to them) stay connected. The reverse operation is also 
considered
as a minimal modification. We will discuss further below this minimal 
modification.
\par  \medskip
\noindent {\bf B - {\it Proof of the independence of the weight-ratio (51) on
$\{s_{j+1}\}$}}\par

We know that for spanning trees in $V_{j+1}$, given two vertices $s_{1j}$ and
$s_{2j}$ of $\{s_j\}$, there are two possibilities : \par

i) they are not connected in $V_j$, \par

ii) they are connected in $V_j$ by a self-avoiding path on the spanning 
tree. \par
\noindent Let us call such a path in $V_j$ $P_1(s_{1j}, s_{2j})$. From such a
path, spanning trees in $V_{j+1}$ can be constructed by rooting branches on 
it, and
then second branches rooted on the first branches, and then again branches 
rooted
on these second branches, and so on until every vertex in $V_{j+1}$ is 
incident
with a branch. Let us suppose for a while that a self-avoiding path 
$P_2(s_{2j},
s_{1j})$ is entirely in $V_{j+1} - V_j$. From this path again spanning trees 
in
$V_{j+1}$ can be constructed in the same way as for $P_1(s_{1j}, s_{2j})$.
However, on these last spanning trees $s_{1j}$ and $s_{2j}$ are disconnected 
in
$V_j$. We remark that the succession of paths $P_1(s_{1j}, s_{2j})$ 
$P_2(s_{2j},
s_{1j})$ forms a loop ${\cal L}$ which crosses the border of $V_j$ with 
$V_{j+1}
- V_j$ at $s_{1j}$ and $s_{2j}$. Then, \par

i) if we cut the loop ${\cal L}$ by removing one propagator $(v_1v_2)$ on
$P_1(s_{1j}, s_{2j})$, $v_1$ and $v_2$ being the end-vertices of the 
propagator,
we get a self-avoiding path on ${\cal L}$, $P(v_2, v_1)$ from which spanning
trees in $V_{j+1}$ can be constructed in which $s_{1j}$ and $s_{2j}$ are not
connected in $V_j$. $P(v_2, v_1)$ is considered as $P_2(s_{2j}, s_{1j})$ with
two branches on $P_1(s_{1j}, s_{2j})$ rooted at $s_{1j}$ and $s_{2j}$. \par

ii) If we cut the loop ${\cal L}$ by removing one propagator $(v'_1v'_2)$ on
$P_2(s_{2j}, s_{1j})$, $v'_1$ and $v'_2$ being the end-vertices of this
propagator, we get a self-avoiding path $P(v'_1, v'_2)$ from which spanning
trees in $V_{j+1}$ can be constructed in which $s_{1j}$ and $s_{2j}$ {\it are
connected in $V_j$}. $P(v'_1, v'_2)$ is considered as $P_1(s_{1j}, s_{2j})$ 
with
two branches on $P_2(s_{2j}, s_{1j})$ rooted at $s_{1j}$ and $s_{2j}$. \par

Cutting $(v_1v_2)$ instead of $(v'_1v'_2)$ on ${\cal L}$ {\it then defines a 
minimal
modification} which modifies $\{s_j\}$. Calculating the weight of the sum of
spanning $m'$-trees in $V_{j+1}$ obtained by cutting $m'-1$ propagators of
the spanning trees constructed in i) we get a weight $W_{m_1}^{m'}(\{s_j\}_1,
\{s_{j+1}\})$ if the cutting is made such as to preserve $\{s_j\}_1$ and
$\{s_{j+1}\}$. Calculating the weight of the sum of spanning $m'$-trees in
$V_{j+1}$ obtained by cutting $m' - 1$ propagators of the spanning trees
constructed in ii) we get a weight $W_{m_2}^{m'}(\{s_j\}_2, \{s_{j+1}\})$ if
again the cutting is made such as to preserve $\{s_j\}_2$ and $\{s_{j+1}\}$. 
We
will see that for a given ${\cal L}$ the ratio the respective contributions
from i) and from ii) to the weights is easily obtained. (Here we have assumed
that $\{s_j\}_1$ and $\{s_j\}_2$ are related by only one minimal 
modification). However, to
obtain the total contribution to the weights we have, of course, to sum over 
all allowed
${\cal L}$. \par

The justification for the use of the loop ${\cal L}$ is that it allows us to 
make
a systematic correspondence between paths $P_2(s_{1j}, s_{2j})$ which have
some part in $V_{j+1} - V_j$ and paths $P_1(s_{1j}, s_{2j})$ which stay in
$V_j$. Let us call $P_2(s_{kj}, s_{k+1j})$ a sub-path of $P_2(s_{1j},
s_{2j})$ in $V_{j+1} - V_j$, $s_{kj}$ and $s_{k+1j}$ being vertices of
$\{s_j\}$. To $P_2(s_{kj}, s_{k+1j})$ we associate a path $P_1(s_{kj},
s_{k+1j})$ in $V_j$ so that $P_1(s_{kj}, s_{k+1j})$ $P_2(s_{k+1j} ,
s_{kj})$ will form a loop ${\cal L}_k$. Let us give an ensemble $\{{\cal
L}_k\}$ of these loops, each being not incident with any another one. 
Relating
the loops ${\cal L}_k$ we have paths $P(s_{kj}, s_{k'j})$ in $V_j$, with $k'
\not= k+1$, which are not incident with $\{ {\cal L}_k\}$ except at $s_{kj}$
and $s_{k'j}$ and which are common to $P_1(s_{1j}, s_{2j})$ and $P_2(s_{1j},
s_{2j})$. It is clear then, that the sets $\{{\cal L}_k\}$ allows us to {\it
make a systematic correspondence between all self-avoiding paths $P_1(s_{1j},
s_{2j})$ and all self-avoiding paths $P_2(s_{1j}, s_{2j})$}~:
$$P_1(s_{1j} , s_{2j}) \ \hbox{$\{{\cal L}_k\} \atop \longleftrightarrow $} 
\ P_2(s_{1j}, s_{2j}) \eqno(52)$$

\noindent where $\{{\cal L}_k\}$ and the paths $P_1(s_{1j}, s_{2j})$,
$P_2(s_{1j}, s_{2j})$ are incident with $\{s_j\}$ at the same vertices. \par

Now, on each ${\cal L}_k$ we can make a cut either in $V_j$ or in $V_{j+1} -
V_j$ in order to go from a partition where $s_{kj}$ and $s_{k+1j}$ are
disconnected in $V_j$ to partition where they are connected in $V_j$. 
However,
we see that for each $k$ the cutting of a loop ${\cal L}_k$ corresponds to a
different minimal modification of the partition $\{s_j\}$. Moreover, each 
loop
${\cal L}_k$ has to be cut in order to have a tree. Therefore, to $p$ loops
${\cal L}_k$ in $\{{\cal L}_k\}$ we have a set of $p$ cuttings. Here, {\it
requiring only one minimal modification to take place}, we want to cut
$P_1(s_{1j}, s_{2j})$ at most once. Suppose this cut takes place on a given
loop ${\cal L}_k$. Then, we will associate to $P_1(s_{1j}, s_{2j})$ the path
$P_2(s_{1j}, s_{2j})$ which differs from $P_1(s_{1j}, s_{2j})$ only on ${\cal
L}_k$, i.e. {\it we will have only one ${\cal L}_k$ in $\{{\cal L}_k\}$}. 
Then,
$P_1(s_{1j}, s_{2j})$ will consist of the succession of paths $P(s_{1j},
s_{kj})$ $P_1(s_{kj}, s_{k+1j})$ $P(s_{k+1j}, s_{2j})$, all in $V_j$, see 
fig. 2. The
simplest topology appears when $s_{1j}$ is $s_{kj}$ and $s_{2j}$ is
$s_{k+1j}$, ${\cal L}_k$ becoming the loop ${\cal L}$ described before. In
the following, we will treat this simple case first because the reasoning is
almost unchanged passing from ${\cal L}$ to ${\cal L}_k$. We now turn to the
construction of spanning $m'$-trees in $V_{j+1}$, and first of spanning trees
in $V_{j+1}$. \par

We observe that the branches in i) and ii) are exactly the same. The
corresponding spanning trees in $V_{j+1}$ only differ in the way ${\cal L}$
is cut. \par

From the spanning trees, $m'$-trees in $V_{j+1}$ are obtained by cutting off
$m' - 1$ propagators either on ${\cal L}$ or on branches. The rule to be
observed is that any sub-tree generated by the cutting should be incident
with at least one vertex of the border of $V_{j+1}$ with $G - V_{j+1}$, i.e.
a vertex of $\{s_{j+1}\}$. Otherwise, a sub-tree would be isolated from all
the others on $G$ and the ensemble of sub-trees could not form a spanning
tree on $G$ as they should. \par

Let us cut branches first. For the reason given above a branch can only be
cut when it is incident or connected to a vertex of $\{s_{j+1}\}$. It is
clear that branches will be cut in exactly the same way for spanning trees in
i) and ii). \par

Then, we come to the eventual cutting of ${\cal L}$, i.e. $P(v_2,v_1)$ or
$P(v'_1, v'_2)$. However, remember that we want that in $\{s_j\}_1$
$P_1(s_{1j}, s_{2j})$ to be cut only once and in $\{s_j\}_2$ not to be cut
at all. Therefore, we don't allow $P(v_2, v_1)$ and $P(v'_1, v'_2)$ to be
cut in $V_j$. However, they can be cut in $V_{j+1} - V_j$. After the
branch-cutting has been completed we focus our attention on those branches
which are still incident or connected with a vertex of $\{s_{j+1}\}$. Let
us call $r_k$ the roots on ${\cal L}$ of these particular branches and
let us order them along ${\cal L}$. It is clear that we can cut ${\cal L}$
only once between $r_k$ and $r_{k+1}$ because otherwise the part of ${\cal
L}$ between two cut propagators, having only branches rooted on it not
incident or connected to $\{s_{j+1}\}$, would be disconnected from all
other sub-trees on $G$, which is forbidden. \par

Moreover, we want the partition $\{s_{j+1}\}$ to be the same for $m'$-trees
constructed from $P(v_2, v_1)$ or $P(v'_1, v'_2)$. Cutting ${\cal L}$
between two roots $r_k$ and $r_{k+1}$ will in general modify $\{s_{j+1}\}$
(except when ${\cal L}$ is cut only once) and therefore any path $P(r_k,
r_{k+1})$ on ${\cal L}$ should be cut or not cut at the same time for the
$m'$-trees generated from $P(v_2, v_1)$ and $P(v'_1, v'_2)$.  \par

The exception is when only one cut is performed on ${\cal L}$, i.e. leaving
$P(v_2, v_1)$ and $P(v'_1, v'_2)$ uncut, because all roots on ${\cal L}$
are still connected and in particular the roots $r_k$. Then, $(v_1, v_2)$
can be anywhere on $P_1(s_{1j}, s_{2j})$ and $(v'_1, v'_2)$ anywhere on
$P_2(s_{2j}, s_{1j})$ in $V_{j+1} - V_j$. \par

Let us return to the general case when ${\cal L}$ is cut more than once.
Cutting ${\cal L}$ at $(v_1v_2)$ or $(v'_1v'_2)$ should not modify
$\{s_{j+1}\}$. Therefore, $(v_1v_2)$ and $(v'_1v'_2)$ should be on the
same path $P(r_{k_1}, r_{k_2})$ on ${\cal L}$, $r_{k_1}$ and $r_{k_2}$
being two consecutive roots of type $r_k$ on ${\cal L}$. Of course,
$P(r_{k_1}, r_{k_2})$ should cross the border of $V_j$ with $V_{j+1} -
V_j$ in order to have $(v_1v_2)$ in $V_j$ and $(v'_1v'_2)$ in $V_{j+1} -
V_j$. \par

Let us now consider roots on ${\cal L}$ of branches in $V_j$ which are
incident with or connected to branches in $V_j$ incident with at least
one vertex of $\{s_j\}$. Let us call such roots $r_b$. Again, it is clear
that ${\cal L}$ can only be cut once between two successive roots $r_b$
and $r_{b+1}$ along ${\cal L}$ for the same reason as for the roots
$r_k$. Furthermore, any cut on a path $P(r_b, r_{b+1})$ on ${\cal L}$
gives rise to $m'$-tree in $V_{j+1}$ with $m$-trees in $V_j$
corresponding to the same partition $\{s_j\}$. Of course, the propagator
$(v_1v_2)$ on ${\cal L}$ is on such a path (which itself is on
$P(r_{k_1}, r_{k_2})$) which we will call $P(r_{b_1}, r_{b_2})$, i.e.
$b_2 = b_1 + 1$, as well as $k_2 = k_1 + 1$. And the propagator
$(v'_1v'_2)$ is on the intersection of $P_2(s_{2j}, s_{1j})$ with
$P(r_{k_1}, r_{k_2})$, this intersection being $P_2(s_{2j}, s_{1j})$ if
$P(r_{k_1}, r_{k_2})$ contains $P_2(s_{2j}, s_{1j})$ (in which case
$r_{k_1}$ and $r_{k_2}$ are in $V_j$), or a path $P(r_{k_1}, s_{1j})$
on $P_2(s_{2j}, s_{1j})$ if $r_{k_1}$ is in $V_{j+1} - V_j$, or a path
$P(s_{2j}, r_{k_2})$ on $P_2(s_{2j}, s_{1j})$ if $r_{k_2}$ is in
$V_{j+1} - V_j$. In any case let us call this intersection path
$P_{int}$ which of course is always in $V_{j+1} - V_j$. Now, {\it
given ${\cal L}$}, we can now write easily the ratio of the weights of
$m'$-trees where $P_{int}$ is cut to those where $P(r_{b_1}, r_{b_2})$
is cut, this is (${\cal L}$ cut more than once, $m' > 1$)
$$\sum_{\ell_2 \subset P_{int}} \ \bar{\alpha}_{\ell_2}/\sum_{\ell_1 \subset
P(r_{b_1}, r_{b_2})} \ \bar{\alpha}_{\ell_1} \eqno(53)$$

\noindent which is a remarkably simple expression. In this minimal
modification $\{s_j\}_1$ goes to $\{s_j\}_2$ and $\{s_{j+1}\}$ is
unchanged. In the case when $m' = 1$, ${\cal L}$ is cut only once, the
roots $r_{k_1}$ and $r_{k_2}$ are irrelevant to determine where $P_2(s_{2j}, 
s_{1j})$ should
be cut in order to keep $\{s_{j+1}\}$ unchanged because all roots on ${\cal 
L}$ are in any
way connected, and $P_{int}$ has to be replaced by $P_2(s_{2j}, s_{1j})$ in 
the above
expression. \par

We now want to make an important observation, i.e. that in the ratio
(53) the only object which may be sensitive to $\{s_{j+1}\}$,
$\{s_j\}$ being fixed, is $P_{int}$ through the position of $r_{k_1}$
or $r_{k_2}$ when $r_{k_1}$ or $r_{k_2}$ are in $V_{j+1} - V_j$.
$P(r_{b_1}, r_{b_2})$ being in $V_j$ is insensitive to
$\{s_{j+1}\}$, because {\it once $\{s_j\}$ is fixed, $r_{b_1}$ and
$r_{b_2}$ cannot depend on the structure of the $m'$-trees outside
$V_j$}. Or, said otherwise, the structure of $m$-trees in $V_j$ only
depend on $\{s_j\}$. \par

In what follows, we are going to show that in spite of that, we {\it
can constrain the building of the successive volumes $V_j$,
$V_{j+1}, \cdots$ in such a way that the ratio of weights (51) does
not depend on a change of $\{s_{j+1}\}$}. Then, if it is finite or
infinite for one given $\{s_{j+1}\}$ it stays so for any
$\{s_{j+1}\}$. As noted before in the discussion of the
weight-ratio (50) this solves immediately our convergence problem
for $R_i^m(\{s_j\})$. \par \medskip

\noindent {\bf Constraint on the construction of the $V_j$'s} \par

Any propagator going out of $V_j$ is relating a vertex of the
border of the deformed sphere $S_j$ to a vertex of the border of the
deformed sphere $S_{j+1}$ if the vertices at the ends of this propagator are 
one in
$V_j$ and the other one in $V_{j+1} - V_j$. $\sq$ \par
\vskip 3 mm
The later provision takes into account the possibility for the
borders of the deformed $S_j$ and $S_{j+1}$ to coincide on some
domain in which case a propagator going out of $S_j$ would also go out of 
$S_{j+1}$. This
constraint is easy to satisfy because we only need to make the radius of 
$S_{j+1}$
sufficiently close to that of $S_j$ in order to obey it. It has the 
following consequence~:
if a loop ${\cal L}$ enters $V_{j+1} - V_j$ at $s_{2j}$ and reenters $V_j$
at $s_{1j}$, then the part of ${\cal L}$ in $V_{j+1} - V_j$ is a
path
$$(s_{2j} \ s_{2j+1}) \ P(s_{2j+1}, s_{1j+1}) (s_{2j+1}, s_{1j})
\eqno(54)$$
\noindent where $(s_{2j} \ s_{2j+1})$ and $(s_{1j+1} \ s_{1j})$
are two propagators relating vertices on the border of $V_j$ and
$V_{j+1}$. Furthermore any vertex of $P(s_{2j+1}, s_{1j+1})$ not
on the border of $V_{j+1}$ is related to vertices of $V_j$
through paths on $P(s_{2j+1}, s_{1j+1})$ going to vertices on the
border of $V_{j+1}$, see fig. 3. \par

An immediate consequence of the structure of ${\cal L}$ in
$V_{j+1} - V_j$ as shown in (54) is that $s_{1j+1}$ is $r_{k_1}$
or $s_{2j+1}$ is $r_{k_2}$ because being in $\{s_{j+1}\}$ and on
${\cal L}$ they are roots of branches incident with
$\{s_{j+1}\}$, these branches being restricted to one vertex.
Then, $P_{int}$ is simply the propagator $(s_{1j+1} \ s_{1j})$ or
the propagator $(s_{2j} \ s_{2j+1})$ depending on which part of
${\cal L}$ we choose $P(r_{k_1}, r_{k_2})$ to be. {\it However,
in any case, $P_{int}$ depends only on ${\cal L}$ and no more on
$\{s_{j+1}\}$, because the vertices $s_{1j+1}$ and $s_{2j+1}$
will not move on ${\cal L}$ as we change $\{s_{j+1}\}$}. Then,
the ratio (53) will depend on ${\cal L}$ and not $\{s_{j+1}\}$
and the ratio of the weights for two partitions $\{s\}_1$ and
$\{s_j\}_2$ related by a minimal modification will be, {\it
summing over all ${\cal L}$ going out of $V_j$ at $s_{2j}$ and
reentering $V_j$ at $s_{1j}$}. (${\cal L}$ cut more than once,
$m' > 1$),
$$W_{m_1-1}^{m'}(\{s_j\}_2, \{s_{j+1}\})/W_{m_1}^{m'}(\{s_j\}_1, \{s_{j+1}
\}) =$$
$$\bar{\alpha}_{P_{int}} \sum_{T^{m_1-1}} T^{m_1-1}(\bar{i}, \{s_j\}_2)/
\sum_{T^{m_1-1}}
\Big ( \sum_{\ell_1 \subset P(r_{b_1}, r_{b_2})} \bar{\alpha}_{\ell_1} \Big )
T^{m_1-1} (\bar{i}, \{s_j\}_2)  \eqno(55)$$

\noindent where $T^{m_1-1}(\bar{i}, \{s_j\}_2)$ is for the weight
of spanning $(m_1 - 1)$-trees in $V_j$ with partition
$\{s_j\}_2$ and $\bar{\alpha}_{P_{int}}$ is for the
$\bar{\alpha}_{\ell}$ of the propagator $(s_{1j+1} \ s_{1j})$ or
$(s_{2j} \ s_{2j+1})$. The sum over the spanning $(m_1 - 1)$-trees
is provided by cutting in all possible ways compatible with $\{s_j\}_2$ $m_1 
- 2$
propagators from all spanning trees in $V_j$, not going through the
propagator $i$, and obtained from all possible paths
$P_2(s_{2j}, s_{1j})$. Looking at the weight structure in (49.b)
we see that the contribution to weights coming from sub-trees in
$V_{j+1} - V_j$, $\sum\limits_{T^n} T^n(\{s'_j\}_1, \{s_{j+1}\})$,
being the same ones for $W_{m_1-1}^{m'}(\{s_j\}_2, \{s_{j+1}\})$
and $W_{m_1}^{m'}(\{s_j\}_1, \{s_{j+1}\})$, and being
factorized, cancels out in (55). The main feature of the
expression on the right-hand side of (55) is that it does not
depend on $\{s_{j+1}\}$, which is what we were looking for.
\par

We now have to establish the same property for the
contribution coming from spanning $m'$-trees in $V_{j+1}$
where {\it ${\cal L}$ is only cut once}. For this case we need
to separate two classes of $m'$-trees \par

a) those $m'$-trees where for $\{s_j\}_2$, ${\cal L}$ is cut
on $P_{int}$, i.e. on the propagator $(s_{2j} \ s_{2j+1})$ or
$(s_{1j+1} \ s_{1j})$. \par

b) those $m'$-trees where for $\{s_j\}_2$, ${\cal L}$ is cut
on the complement of $P_{int}$ on $P_2(s_{2j}, s_{1j})$, which we
will call $P_{int}^{comp}$, but uncut on $P_{int}$. \par

For the $m'$-trees in a) the reasoning is the same as for the
$m'$-trees where ${\cal L}$ is cut more than once and the result (55) is 
valid
for them too. For the $m'$-trees in b), changing
$\{s_{j+1}\}$, {\it we cannot obtain $m'$-trees in which ${\cal
L}$ is cut more than once}, because $P(r_k, r_{k_2})$ would
stay uncut for $\{s_j\}_2$ - corresponding $m'$-trees, which
is forbidden (i.e., for ${\cal L}$ cut more than once,
$P(r_{k_1}, r_{k_2})$ has to be cut for $\{s_j\}_1$ in $V_j$
and for $\{s_j\}_2$ in $V_{j+1} - V_j$, i.e. on $P_{int}$, in order not to 
change
$\{s_{j+1}\}$ passing from $\{s_j\}_1$ to $\{s_j\}_2$). We
remind the reader that when ${\cal L}$ is cut only once all
roots $r_k$ on ${\cal L}$ are connected. Then, for the
$m'$-trees in b) changing $\{s_{j+1}\}$ always keeps all
roots $r_k$ on ${\cal L}$ connected because ${\cal L}$ stays
cut only once. Let us then take {\it all the loops ${\cal L}$
with the same $P_{int}^{comp}$}, for these the contribution of
the weights will be such that (${\cal L}$ cut once, $m' \geq
1$)
$$W_{m_1-1}^{m'}(\{s_j\}_2,
\{s_{j+1}\})/W_{m_1}^{m'}(\{s_j\}_1, \{s_{j+1}\}) =$$
$$\Big ( \sum_{\ell_2 \subset P_{int}^{comp}}
\bar{\alpha}_{\ell_2} \Big ) \sum_{T^{m_1-1}} T^{m_1-1}(\bar{i},
\{s_j\}_2)/\sum_{T^{m_1-1}} \Big ( \sum_{\ell_1 \subset
P(r_{b_1}, r_{b_2})} \bar{\alpha}_{\ell_1} \Big )
T^{m_1-1} (\bar{i} , \{s_j\}_2) \eqno(56)$$

\noindent which, again, is independent of $\{s_{j+1}\}$. \par

Now, because each contribution to the weight
$W_{m_1}^{m'}(\{s_j\}_1, \{s_{j+1}\})$ is multiplied by a
factor independent of $\{s_{j+1}\}$ when the corresponding
contribution to $W_{m_1-1}^{m'}(\{s_j\}_2, \{s_{j+1}\})$ is
taken, we have the following theorem~: \par
 
\noindent {\bf Theorem 2a} \par 
Let us consider a minimal modification of $\{s_j\}$,
$\{s_j\}_1 \to \{s_j\}_2$ where $\{s_{j+1}\}$ remains
unchanged. In $\{s_j\}_2$, $s_{1j}$ and $s_{2j}$, vertices
of $\{s_j\}$ are connected by a self-avoiding path
$P_1(s_{1j}, s_{2j})$ in $V_j$. In $\{s_j\}_1$, $P_1(s_{1j},
s_{2j})$ is cut once. The $m'$-trees in $V_{j+1}$
corresponding to $\{s_j\}_2$ are obtained from $P_1(s_{1j},
s_{2j})$. The $m'$-trees in $V_{j+1}$ corresponding to
$\{s_j\}_1$ are obtained from a self-avoiding path
$P_2(s_{2j}, s_{1j})$ with all propagators in $V_{j+1} -
V_j$, its only vertices on the border of $V_{j+1} - V_j$ 
with $V_j$ being $s_{1j}$ and $s_{2j}$. \par

Then, the ratio of the sum of weights of $m'$-trees
corresponding to $\{s_j\}_2$ to the sum of the weights of
$m'$-trees corresponding to $\{s_j\}_1$ is independent of
$\{s_{j+1}\}$. $\sq$ \par
\vskip 3 mm

We have examined so far the simplest case where the loop
${\cal L}_k$ was taken as the loop ${\cal L}$ going out of
$V_j$ at $s_{2j}$ and in $V_j$ at $s_{1j}$. In general, as
discussed earlier, a path $P_2(s_{1j}, s_{2j})$ may have
some parts $P_2(s_{kj}, s_{k+1j})$ in $V_{j+1} - V_j$.
Confining ourselves to the minimal modification of
$\{s_j\}$, only one such $P_2(s_{kj}, s_{k+1j})$ or one
${\cal L}_k$ is relevant. \par

We can repeat the reasoning followed with the loop ${\cal
L}$ for the loop ${\cal L}_k$, the only change being that
two self-avoiding paths in $V_j$, $P(s_{1j}, s_{kj})$ and
$P(s_{k+1j}, s_{2j})$ will be rooted at $s_{kj}$ and
$s_{k+1j}$ respectively on ${\cal L}_k$. Then, we have
to divide the paths $P_1(s_{1j}, s_{2j})$ into two classes
in order to avoid a double-counting~: \par

a) those which are incident with a vertex of $\{s_j\}$
other than $s_{1j}$ and $s_{2j}$ only once \par

b) those which are incident with at least two vertices
$s_{kj}$ and $s_{k'j}$ of $\{s_j\}$. \par

The paths a) will be associated to the loop ${\cal L}$. The paths b) will be
associated to a loop ${\cal L}_k$ with $s_{k'j}$ being then
noted $s_{k+1j}$. Of course, in this case $P_1(s_{1j},
s_{2j})$ can be incident with other vertices of $\{s_j\}$
as well. We only need to exhaust all pairs of vertices
$s_{kj}$, $s_{k'j}$ in order to form all possible loops
${\cal L}_k$. In this way, all different paths
$P_1(s_{1j}, s_{2j})$ are taken into account only once.
Theorem 2a applied to each pair $s_{kj}$, $s_{k'j}$
replacing $s_{1j}$ and $s_{2j}$ then provides the
independence of the weight ratio (51) on $\{s_{j+1}\}$.
\par

The only thing which is left to prove is that, indeed, any
partition $\{s_j\}_2$ can be obtained from another one
$\{s_j\}_1$ by a series of minimal modifications. So let
us consider the res\-pec\-tive situation of two vertices
$s_{1j}$ and $s_{2j}$ in $\{s_j\}_1$ and in $\{s_j\}_2$.
Suppose that they are unconnected in $\{s_j\}_1$ and
connected in $\{s_j\}_2$. Then, it is easy to see that a
minimal modification will allow to disconnect them,
passing from $\{s_j\}_2$ to $\{s_j\}_3$, leaving
weight-ratios corresponding to those in (51) independent
of $\{s_{j+1}\}$. Then, we will consider the respective
situation of any two other $s_j$'s and repeat the
operation until we obtain $\{s_j\}_1$, having always
weight-ratios insensitive to $\{s_{j+1}\}$. \par

Let us make also a remark about the topology of $G$ in
$V_{j+1}$. $G$ can be disconnected into several pieces in
$V_{j+1}$, although $G$ itself is taken to be connected
and even 1-line and 1-vertex irreducible. When $G$ is
multiply connected in $V_{j+1}$ we consider each
connected piece separately for the construction of
spanning trees on them and the eventual cutting of
propagators. The resulting weight for $m'$-trees will
simply be the product of the weights of all connected
pieces of $G$ in $V_{j+1}$. Of course, in this case $m' >
1$ but the above reasoning is essentially unchanged. We
then are able to write the following theorem~:\par
 
\noindent {\bf Theorem 2b} \par 
For any two different partition $\{s_j\}_1$ and
$\{s_j\}_2$ of vertices $s_j$ on the border of $V_j$ with
$V_{j+1} - V_j$ the ratio of weights (51) is independent of
$\{s_{j+1}\}$ if we impose the constraint on the
construction of the $V_j$'s described earlier and which
can always be satisfied. It follows that the ratio (50) 
$$\sum_{\{s_j\}_M} W_{m_M}^{m'}\left ( \{s_j\}_M, \{s_{j+1}\}
\right )/\sum_{\{s_j\}} W_{m}^{m'}\left ( \{s_j\}, \{s_{j+1}\}
\right )$$

\noindent is also independent of $\{s_{j+1}\}$. Therefore,
when this ratio is infinite it is so for all $\{s_{j+1}\}$
and $R_i^{m'}(\{s_{j+1}\})$ takes a unique value
$R_i^{m}(\{s_{j+1}\}_M)$. When this ratio is finite, it
also stays finite for all $\{s_{j+1}\}$ and therefore the
convergence condition (41) is satisfied. $\sq$ \par
\vskip 3 mm

In order to have $R_i^{m}(\{s_{j}\})$ to converge as $j
\to \infty$, we also need to have this ratio finite for
some finite value of $j$. This condition is naturally
satisfied if we impose the constraint that $V_1$ should
only contain a finite number of propagators. This
condition, of course, can always be satisfied by choosing
a sufficiently small radius for the sphere $S_1$. \par

\medskip 
\noindent {\bf C - {\it Extension of the convergence proof to trees}}\par 

We now want to extend the inequality (36) of theorem 1 to the spanning
trees in $V_j$ and $V_{j+1}$. For that purpose let us define a (spanning) 
multiple
tree (or $m'$-tree) $T^{m'}(i, \lbrace s_{j+1} \rbrace)$ contained in 
$V_{j+1}$, one
component of which is going through $i$. $\lbrace s_{j+1} \rbrace$ stands 
for a
partition of all border-vertices of $V_{j+1}$ with $G - V_{j+1}$ with which 
$T^{m'}(i, \lbrace
s_{j+1} \rbrace)$ is incident. As for paths, $T^{m'}(\bar i, \lbrace s_{j+1} 
\rbrace)$ has
the same properties as $T^{m'}(i, \lbrace s_{j+1} \rbrace)$,  except that it 
does not
go through $i$ but is still incident with $v_i$ through one to its 
component-tree. In
the following, in order to have a simpler notation we will imply that {\it a 
summation
over all \ $T$'s of the kind is made whenever the symbol $T$ appears}. Then, 
we can
write, $\lbrace s_j \rbrace$ being some partition of vertices of the border 
of $V_j$
with $V_{j+1} - V_j$ and $\lbrace s_{j+1} \rbrace$ a corresponding partition 
for $V_{j+1}$,
($T^m(i, \lbrace s_j \rbrace$) is a $m$-tree contained in $V_j$, $T(\lbrace 
s_j^t \rbrace$) is
a tree in $V_{j+1} - V_j$ with no vertex on the border of $V_{j+1}$ and 
incident with
the border of $V_j$ at $\lbrace s_j^t \rbrace$, $T(\lbrace s_j^u \rbrace \, 
\lbrace
s_{j+1}^u \rbrace$) is a tree in $V_{j+1} - V_j$ incident with the borders 
of $V_j$
and $V_{j+1}$ at $\lbrace s_j^u \rbrace$ and  $\lbrace s_{j+1}^u \rbrace$
respectively, and $T(\lbrace s_{j+1}^v \rbrace)$ is a tree in $V_{j+1} - 
V_j$ not
incident with the border of $V_j$ but incident with the border of $V_{j+1}$ 
at
$\lbrace s_{j+1}^v \rbrace)$  $$T^{m'}(i, \lbrace s_{j+1} \rbrace) = 
\sum_{m=1}^{M_j}
\sum_{\lbrace s_j \rbrace} T^m(i, \lbrace s_j \rbrace).$$ $$\sum_{\lbrace 
\lbrace
s_j^t \rbrace \rbrace} \prod_{t=1}^x T(\lbrace s_j^t \rbrace) \sum_{\lbrace 
\lbrace
s_j^u \rbrace , \lbrace s_{j+1}^u \rbrace \rbrace} \prod_{u=1}^y T(\lbrace 
s_j^u
\rbrace, \lbrace s_{j+1}^u \rbrace) \sum_{\lbrace \lbrace s_{j+1}^v \rbrace
\rbrace} \prod_{v=1}^z T(\lbrace s_{j+1}^v \rbrace) \eqno(57)$$

\noindent ($\{ \{s_j^u \}$, $\{s_{j+1}^u \} \}$ is the ensemble of 
border-vertices
of the $y$ trees $T(\{s_j^u \}$,~$\{s_{j+1}^u \})$, $\{ \{ s^t \} \}$ is the
ensemble of the border-vertices of the $x$ trees $T(\{ s_j^t \})$ and $\{ \{
s_{j+1}^v \} \}$ is the ensemble of border-vertices of the $z$ trees $T(\{
s_{j+1}^v \})$) for $x$, $y$, $z$ $\geq$ 1 and with $$y + z \geq m' \geq z
\eqno(58)$$

\noindent When $x$ = 0, $y$ = 0 or $z$ = 0 the corresponding product in (57) 
is
equal to one (no tree). The relation (51) can be explained by saying that $y 
+
z$ is the maximum number of components of $T^{m'}(i, \lbrace s_{j+1} 
\rbrace)$
and $z$ the minimum number of its components. A tree $T(\lbrace s_j^t
\rbrace)$ in $V_{j+1} - V_j$ should be connected to at least one tree in 
$V_j$,
giving 
$$\forall \ t, \ \ \ \ \ \lbrace s_j \rbrace \cap \lbrace s_j^t \rbrace
\not= \phi \eqno(59)$$

\noindent  If $\lbrace s_j^{m_c} \rbrace$, ($m_c = 1, \cdots , m)$, denotes 
the set of
border-vertices of $V_j$ of one component-tree of $T^m(i, \lbrace s_j 
\rbrace)$,
i.e. if $\lbrace s_j \rbrace$ = $\lbrace \lbrace s_j^{m_c} \rbrace \rbrace$, 
then,
avoiding loops, 
$$\lbrace s_j^t \rbrace \cap \lbrace s_j^{m_c} \rbrace = {\rm  at \
most \ one} \ s_j \eqno(60.a)$$

$$\lbrace s_j^u \rbrace \cap \lbrace s_j^{m_c} \rbrace = {\rm at \ most 
\ one} \ s_j
\eqno(60.b)$$

\noindent Any component-tree of $T^m(i, \lbrace s_j \rbrace)$ should be 
connected
to at least one tree in $V_{j+1} - V_j$ giving 
$$\forall \ m, \ \ \ \lbrack \lbrace
\lbrace s_j^t \rbrace \rbrace \cup \lbrace \lbrace s_j^u \rbrace \rbrace 
\rbrack
\cap \lbrace s_j^{m_c} \rbrace \not= \phi \eqno(60.c)$$

The relations (60) insure that a component-tree of $T^m(i, \lbrace s_j
\rbrace)$ in $V_j$ should be incident with at least one component tree
in $V_{j+1} - V_j$ through at most one vertex. Indeed, (59) and (60)
fix the topology of contact between the component-trees in $V_j$ and
$V_{j+1} - V_j$ in order to leave no component-tree isolated and in a
way that avoids the formation of any loop. Let us denote $\lbrace
\lbrace s_j^t \rbrace \rbrace \cup \lbrace \lbrace s_j^u \rbrace
\rbrace$ by $\lbrace s'_j \rbrace$ which will be the ensemble of the
vertices on the border of $V_j$ of the trees in $V_{j+1} - V_j$. Then,
it is clear that once $\lbrace s_j \rbrace$ and $\lbrace s'_j \rbrace$
have been fixed, the summation over trees in $V_j$ and $V_{j+1} - V_j$
is factorizable because no interaction occurs between the trees in
$V_j$ and those in $V_{j+1} - V_j$ (apart from their contact at the
common border of $V_j$ and $V_{j+1} - V_j$). Note even that once $\lbrace s_j
\rbrace$ is fixed the summation over all $T^m(i, \lbrace s_j \rbrace )$ (or
over all $T^m(\bar{i}, \lbrace s_j \rbrace)$) does not depend on $\lbrace 
s'_j
\rbrace$. Consequently, the sum over all $n$-trees in $V_{j+1} - V_j$, {\it
including the sum over $\lbrace s'_j \rbrace$}, can be factorized out. \par
Again, in a way analogous to that of paths let us define 
$$R_i^m(\lbrace s_j \rbrace) = \sum_{T^m} T^m(i, \lbrace s_j
\rbrace)/\sum_{T^m} T^m(\bar i, \lbrace s_j \rbrace) \eqno(61)$$

\noindent for which we demonstrate as for paths the following \par
 \noindent {\bf Theorem 3} \par 
$$R_i^m(\lbrace s_j \rbrace)_{min} < R_i^{m'}(\lbrace s_{j+1} \rbrace) < 
R_i^m(\lbrace
s_j \rbrace)_{max} \eqno(62)$$

\noindent where $R_i^m(\lbrace s_j \rbrace)_{min}$ and $R_i^m(\lbrace s_j
\rbrace )_{max}$ are respectively the minimum and the maximum value of
$R_i^m(\lbrace s_j \rbrace)$ for all $m$'s or $R_i^{m'}(\{s_{j+1}\})$ has 
converged
to either $R_i^m(\{s_j\})_{min}$ or $R_i^m(\{s_j\})_{max}$. \par  

\noindent {\bf Proof} \par 

It is clear, using (61) in (57) that $R_i^{m'}(\lbrace
s_{j+1} \rbrace)$ is a mean-value of $R_i^m(\lbrace s_j \rbrace)$ considered 
as a
function of $m$, $\lbrace s_j \rbrace$ and $\lbrace \lbrace s_j \rbrace
\rbrace$, $\lbrace \lbrace s'_j \rbrace \rbrace$. However, due to the
factorization property mentioned above (once $\lbrace s_j \rbrace$ is fixed,
$\lbrace s'_j \rbrace$ does not have an influence over the $m$-trees in 
$V_j$),
the functional dependence of $R_i^m(\lbrace s_j \rbrace)$ is indeed 
restricted
to $m$ and $\lbrace s_j \rbrace$. Moreover, theorem 2b either excludes the 
limiting
cases where the inequalities become equalities or makes $R_i^{m'}(\{s_{j+1}
\})$
equal to $R_i^m(\{s_j\})_{min}$ or $R_i^m(\{s_j\})_{max}$ for any $\{s_{j+1}
\}$. It
therefore follows that either (62) is true or $R_i^{m'}(\{s_{j+1}\})$ has 
converged.
$\sq$ \par 

\noindent \underbar{Remark} \par 

We note that $R_i^m(\lbrace s_j \rbrace)$ (and $T^m(i, \lbrace s_j \rbrace$) 
is a
function of a partition $\lbrace s_j \rbrace$ which covers any part of the 
border of
$V_j$ on the surface of the deformed $S_j$ independently of the partition $
\lbrace
s_{j+1} \rbrace$ of $T^{m \prime}(i, \lbrace s_{j+1} \rbrace)$. Therefore, 
the sum
$\displaystyle {\sum_{\lbrace s_j \rbrace}}$ in (57) is a sum over the {\it 
whole}
part of border of $V_j$ on the deformed surface $S_j$. This is in contrast 
with the
corresponding sums $\displaystyle {\sum_{s_j^0}}$ or $\displaystyle {\sum_{
\lbrace
s_j^l \rbrace}}$ in (34) for paths which may cover only part of the border 
of $V_j$
on the deformed surface $S_j$, depending on $s_{j+1}$ in $P(i, s_{j+1})$. 
Therefore,
in the case of trees, the topology of $V_{j+1} - V_j$ does not intervene to 
possibly
limit the range of $\lbrace s_j \rbrace$. Hence, we have a unique value for
$R_i^m(\lbrace s_j \rbrace)$ and this solves problem b) of the preceding 
section. The
repeated use of (62) will make all $R_i^m(\lbrace s_j \rbrace)$ converge 
towards the
same value $R_i^{\infty}$. \par

\medskip \noindent {\bf 5. The factorization of trees on G} \medskip 

We now proceed to the proof of the relation (24), the crucial factorization 
property
of spanning trees on $G$. \par 

Let us consider a volume $V_j$ with $j \to \infty$, and a partition of its
border-vertices $\lbrace s_j \rbrace$ and let us denote by $R_i^{\infty}$ 
the common
value to which all $R_i^m(\lbrace s_j \rbrace)$ tend as $j \to \infty$. 
Recalling
(16) and (17) we have, $T^n$ being an $n$-tree on $G - V_j$, (a sum over $m$ 
is implied in
the sum over all $\{s_j\}$ and a sum over $n$ is implied in the sum over all 
$\{s'_j\}$
compatible with $\{s_j\}$)  $$\bar a_i = \Delta (\bar \alpha) \sum_{{\cal T} 
\supset i} \
\prod_{l \in {\cal T}} \bar \alpha_l^{-1}$$ $$= \Delta (\bar \alpha) \sum_{
\lbrace s_j
\rbrace , T^m} T^m (i, \lbrace s_j \rbrace) \sum_{\lbrace s'_j \rbrace , 
T^n} T^n (\lbrace
s'_j \rbrace) \eqno(63.a)$$

$$\bar b_i = \bar \alpha_i^{-1} \Delta (\bar \alpha) \sum_{{\cal T}
\not\supset i} \ \prod_{l \in {\cal T}} \bar \alpha_l^{-1}$$ $$= \bar
\alpha_i^{-1} \Delta (\bar \alpha) \sum_{\lbrace s_j
\rbrace , T^m} T^m (\bar{i}\lbrace s_j \rbrace) \sum_{\lbrace s'_j
\rbrace , T^n} T^n (\lbrace s'_j \rbrace) \eqno(63.b)$$

\noindent $\lbrace s'_j \rbrace$ is a subset of the vertices of $V_j$ 
belonging to
$T^n(\lbrace s'_j \rbrace)$, and $T^n \cup T^m$ form a spanning tree of $G$.
Notice that the sets $\lbrace s'_j \rbrace$ and $\lbrace s_j \rbrace$ are
in general different, a branch of $T^m$ can end at one $s_j$ without being
incident at that $s_j$ with one branch of $T^n$. For any given
$m$, $\lbrace s_j \rbrace$,  $T^m(i, \lbrace s_j \rbrace)$ can be replaced by
$R_i^{\infty} \ T^m(\bar i, \lbrace s_j \rbrace)$ and we therefore get 
$$\bar a_i/ \bar b_i = R_i^{\infty} \bar \alpha_i \ \ \ . \eqno(64)$$

Remember that by a given $\lbrace s_j \rbrace$ we mean a given $\lbrace
\lbrace s_j^{m_c} \rbrace \rbrace$ where $\lbrace s_j^{m_c} \rbrace$ is a 
set of
$s_j$'s belonging to the same component-tree of $T^m$. In the same way
$\lbrace s'_j \rbrace$ means $\lbrace \lbrace s_j^{'n_c} \rbrace \rbrace$
where $\lbrace s_j^{'n_c} \rbrace$ belongs to one component-tree of $T_n$.
\par 

Looking now at (22) and (23) we can write, using the same notations, 
$$\bar d_{i,k} = \sum_{\lbrace s_j \rbrace ,T^m} s_{{\cal C}_k}\ \nu^{-1} ({
\cal
C}_k) \ T^m (i, \lbrace s_j \rbrace) \sum_{\lbrace s'_j \rbrace ,T_k^n} T_k^n
(\lbrace s'_j \rbrace) \eqno(65.a)$$

$$\bar e_{i, k} = \bar \alpha_i^{-1} \sum_{\lbrace s_j \rbrace , T^m} s_{{
\cal C}_k} \
\nu^{-1} ({\cal C}_k^i) \ T^m (\bar{i}, \lbrace s_j \rbrace) \sum_{\lbrace 
s'_j
\rbrace , T_k^n} T_k^n(\lbrace s'_j \rbrace) \eqno(65.b)$$

\noindent where $T_k^n$ is a $n$-tree belonging to $G - V_j$ and going 
through
the propagator $k$ which is assumed to be outside $V_j$. \par

We see that, compared to (63), the expressions (65) have exactly the same 
structure
except for the factors $s_{{\cal C}_k} \ \nu^{-1}({{\cal C}_k})$ and $s_{{
\cal C}_k} \
\nu^{-1}({\cal C}_k^i)$. We now want to demonstrate that, indeed,
$$\lbrack s_{{\cal C}_k} \ \nu^{-1}({{\cal C}_k}) \rbrack / \lbrack s_{{\cal 
C}_k^i} \
\nu^{-1} ({{\cal C}_k}^i ) \rbrack = 1 + \varepsilon \eqno(66)$$

\noindent where $\varepsilon$ is infinitesimal. Let us recall that $\nu({\cal
C}_k)$ counts the number of pro\-pa\-ga\-tors on the surface ${\cal S}_k({
\cal C}_k)$
cutting $G$ into two disjoint pieces $G_1({\cal C}_k)$ and $G_2({\cal 
C}_k)$, the same
being true for $\nu({\cal C}_k^i)$ replacing ${\cal C}_k$ by ${\cal C}_k^i$. 
\par

Let us take $j = 1$ for the expressions (65) in order to have a finite 
number of
propagator in $V_1$. First, it is obvious that if ${\cal S}_k$ does not cut 
through
$V_1$ it will cut only the $T^n$'s which are the same $n$-trees in (65.a) and
(65.b) for a given $\{s_j\}$ and we will have $\varepsilon = 0$ in (66). 
Now, if
$j$ grows, ${\cal S}_k$ will remain identical in (65.a) and (65.b) because ${
\cal
C}_k$ and ${\cal C}_k^i$ do not depend on $j$, i.e. on the decomposition of a
spanning tree in $G$ into a $m$-tree in $V_j$ and a $n$-tree in $G - V_j$. 
So,
$\varepsilon$ will remain equal to zero as $j \to \infty$. \par

The non-trivial case occurs when ${\cal S}_k$ cuts through $V_1$. Then, it 
could be
that when $j \to \infty$ the number of propagators cut by ${\cal S}_k$ in 
$V_j$
remains finite. This could arise when there are domains in the deformed 
sphere
$S_j$ which are empty of propagators and of a size large enough so that when 
${\cal
S}_k$ cuts through them it will contain only a finite number of propagators 
in
$V_j$. An example of this situation is provided when $G$ has a topology such 
that
it consists of infinite ladders (which may join and separate themselves 
creating a
sort of effective field theory of Reggeons). \par

To study the situation where $S_k$ cuts through $V_1$ let us consider, for a 
given spanning
tree ${\cal T}$ on $G$, the sum
$$\Sigma_{\cal T}(G) = \sum_{k \in {\cal T}} \bar{\alpha}_k \ s_{{\cal C}_k}
\ \nu^{-1} ({\cal C}_k) \eqno(67)$$

\noindent and a similar sum for the part of ${\cal T}$ in $V_1$, ${\cal T}_1$
$$\eqalignno{
\Sigma_{\cal T}(V_1) &= \sum_{k \in {\cal T}_1} \bar{\alpha}_k \ s_{{\cal
C}_k} \nu^{-1}({\cal C}_k) \cr
&=< \bar{\alpha}_k \ s_{{\cal C}_k}>_{{\cal T}_1} \sum_{k \in {\cal T}_1} 
\nu^{-1}({\cal
C}_k) &(68) \cr }$$

\noindent where $<\bar{\alpha}_k \ s_{{\cal C}_k}>_{{\cal T}_1}$ is the 
mean-value of
$\bar{\alpha}_k \ s_{{\cal C}_k}$ for all propagators $k$ in $V_1$ belonging 
to ${\cal
T}$. The maximum value of $\sum_{\cal T}(V_1)$ is obtained when $\nu ({\cal 
C}_k)$ for
all $k$'s belonging to ${\cal T}_1$ is constant, meaning that the 
corresponding $S_k$'s
each contain a finite number of propagators. Then, $\sum_{\cal T}(V_1)$ is 
equal to
$<\bar{\alpha}_k \ s_{{\cal C}_k}>_{{\cal T}_1}$ multiplied by some 
constant. \par

We can express $\sum_{\cal T}(G)$ in the same way, writing
$$\Sigma_{\cal T}(G) = <\bar{\alpha}_k \ s_{{\cal C}_k}>_{\cal T} \sum_{k \in
{\cal T}}  \nu^{-1}({\cal C}_k) \eqno(69)$$

\noindent where $<\bar{\alpha}_k \ s_{{\cal C}_k}>_{\cal T}$ is the 
mean-value of $\bar{\alpha}_k \ s_{{\cal
C}_k}$ over ${\cal T}$. Now, we can calculate a lower bound for
$\displaystyle{\sum_{k \in {\cal T}}} \nu^{-1}({\cal C}_k)$. \par

This comes from the fact that every propagator in $S_k$ has to be incident 
with a vertex of
the sub-trees in $G_1({\cal C}_k)$ and $G_2({\cal C}_k)$, the parts of $G$ 
separated by
$S_k$. If a branch consisting of $N$ vertices (and $N - 1$ proapagators) is 
separated in
a $\phi^n$ field theory, the number of propagators cut is $(n - 2)N + 1$ 
which
represents {\it the maximum number of propagators cut for a tree with $N$ 
vertices} (and
of course $N - 1$ propagators). The number of propagators in $G_1({\cal 
C}_k)$, for
example can be taken to vary from zero to $I - L$. Thus, the following 
inequality follows
$$\sum_{k \in {\cal T}} \nu^{-1}({\cal C}_k) > {1 \over n - 2} 
\int_1^{I-L+1} dN/(N+1) =
(1 /(n - 2)) {\rm Log}(I - L + 1) \eqno(70)$$

\noindent with $(1/(n - 2)) {\rm Log}(I - L + 1)$ tending to infinity as $I 
\to
\infty$. Therefore, if the ratio 
$$<\bar{\alpha}_k \ s_{{\cal C}_k}>_{{\cal T}_1}/<\bar{\alpha}_k \ s_{{\cal 
C}_k}>_{\cal T}
\eqno(71)$$

\noindent stays finite, the ratio
$$\Sigma_{\cal T}(G)/\Sigma_{\cal T}(V_1) \eqno(72)$$

\noindent is infinite as $I \to \infty$ and the contribution to (65.c) and 
(65.b) coming
from $S_k$'s cutting through $V_1$ can be neglected. It follows, then, that 
(66) will be
true. \par

Let us now remark that (71) may be infinite in the case where some internal 
lines incident
with vertices in $V_1$ carry a momentum infinite with respect to the momenta 
of external
lines incident with $G - V_1$. Then, $S_k$ may cut $V_1$ in such a way as to 
separate such
lines, giving an infinite $s_{{\cal C}_k}$, while a cancellation occurs 
between infinite
momenta when $S_k$ does not cut through $V_1$. Note however that if, 
although infinite, (71)
is equal to 
$$\varepsilon_1 {\rm Log} (I - L) \eqno(73)$$

\noindent with $\varepsilon_1 \to 0$ as $I \to \infty$, the conclusion 
reached above, i.e.
that $\sum_{\cal T}(V_1)$ can be neglected in front of $\sum_{\cal T}(G)$, 
is still valid.
\par

Finally, considering the ratio 
$$\Sigma_{\cal T}(G)/\Sigma_{\cal T}(V_j) \eqno(74)$$

\noindent where $\sum_{\cal T}(V_j)$ is the sum of $\bar{\alpha}_k \ s_{{
\cal C}_k} \nu^{-1}({\cal
C}_k)$ for $k$'s belonging to the part of ${\cal T}$ in $V_j$, we see that 
we have the
inequality 
$$\Sigma_{\cal T}(G)/\Sigma_{\cal T}(V_j) > \left [ (n - 2)^{-1} {\rm Log}(I 
- L)/(I -
L)_{V_j} \right ] <\bar{\alpha}_k \ s_{{\cal C}_k}>_{\cal T}/<\bar{\alpha}_k 
\
s_{{\cal C}_k}>_{{\cal T}_j} \eqno(75)$$

\noindent if $(I - L)_{V_j}$ is the number of lines a spanning tree in $V_j$ 
(a $m$-tree
in $V_j$ has even less lines than $(I - L)_{V_j}$) and $<\bar{\alpha}_k 
\ s_{{\cal
C}_k}>_{\cal T}$ the mean-value of $\bar{\alpha}_k \ s_{{\cal C}_k}$ over ${
\cal T}_j$,
the $m$-tree part of ${\cal T}$ on $V_j$. \par

Provided the ratio on the right of (75) is infinite we can neglect in (20) 
and (21) the
sum over $k$, $k$ belonging to $V_j$, even as $(I - L)_{V_j}$ goes to 
infinity, as was
claimed in section~3. \par
 
Then, $T^m(i, \{s_j \})$ can be replaced by $R_i^{\infty} \ T^m(\bar{i}, \{ 
s_j \})$
in (65.a) and (65.b) as in (63.a) and (63.b) with the result
$$\bar{d}_{i,k}/\bar{e}_{i,k} = R_i^{\infty} \bar{\alpha}_0 \eqno(76)$$

\noindent which was sought after. This entails that (24) and thereby (23) are
verified. As said in section 2 this, in turn, makes $Q_G(P_v, \{\bar{
\alpha}_i\})$
insensitive to the replacement $\bar{\alpha}_i \to \bar{\infty}$ and a unique
$\bar{\alpha}$ can be used to evaluate $F_G$ in a super-renormalizable scalar
field theory. \par \medskip  

\noindent {\bf 6. Conclusion} \medskip \par 

Our initial aim was to put the Gaussian representation for propagators on a 
firm
footing. Using the well-known $\alpha$-representation we are able to prove 
that the
parameter $\alpha$ which measures the inverse of the variance of that 
Gaussian can be
taken everywhere equal to some unique $\bar \alpha = O(1/I)$ where $I$ is 
the number
of internal lines of a Feynman graph $G$. We did this for a 
super-renormalizable
scalar field theory, although we expect the same result to hold for 
renormalizable
theories as well. But, what is more interesting even, is that we were 
obliged during
the derivation to prove a factorization property of spanning trees on $G$, 
i.e., we
can sum over all graphs in a volume $V_j$, and if $j \to \infty$, the 
structure of
the trees outside $V_j$ is independent of the structure of the same trees in 
$V_1$, a
subpart of $V_j$. We can however imagine that $V_j$ itself is an 
infinitesimal volume
relative to the whole volume of $G$. Then, we can interpret our 
factorization of
trees as the factorization of local sums defined on trees. In other words, {
\it trees
on a Feynman graph can be} {\it used to define a functional
integral}. In fact, we assumed such a functional property in our first
attempt [7] to derive the relation (13). \par

\vfill\supereject
\centerline{\bf References} \bigskip
 \item{[1]} H. B. Nielsen and P. Olesen,
Phys. Lett. $\underline{B32}$ (1970) 203. D. B. Fairlie and H. B. Nielsen, 
Nucl.
Phys. $\underline{B20}$ (1970) 637.
\item{[2]} V. A. Kazakov, I. K.
Kostov and A. A. Migdal, Phys. Lett. $\underline{B157}$ (1985) 295. D. V.
Boulatov, V. A. Kazakov, I. K. Kostov and A. A. Migdal, Phys. Lett.
$\underline{B174}$ (1986) 87.
\item{[3]} F. David, Nucl. Phys.
$\underline{}$ (1985) 543.
\item{[4]} R. Hong Tuan, Phys. Lett.
$\underline{B173}$ (1986) 279 ; Phys. Lett. $\underline{B185}$ (1987) 385 ;
Phys. Lett. $\underline{B244}$ (1990) 413 ; preprint LPTHE 89/15, Orsay
(1989).\item{[5]} M. C. Berg$\grave e$re and J. B. Zuber, Comm. Math.
Phys. $\underline{35}$ (1974) 113. M. C. Berg\`ere and Y. M. P. Lam, J.
M. Phys. $\underline{17}$ (1976) 1546.
\item{[6]} Y. Nambu,
Proceedings International Conference on Symmetries and Quark Models,
Detroit 1969 (Gordon and Breach, N.Y. 1970) p. 269.
\item{[7]} R. Hong
Tuan, Phys. Lett. $\underline{B286}$ (1992) 315.
\item{[8]} C.
Itzykson and B. Zuber, Quantum Field Theory (Mc Graw Hill, New York, 1980)
p. 294. 
\vfill\supereject
\centerline{\bf Figure Captions} \bigskip
{\parindent=1 cm
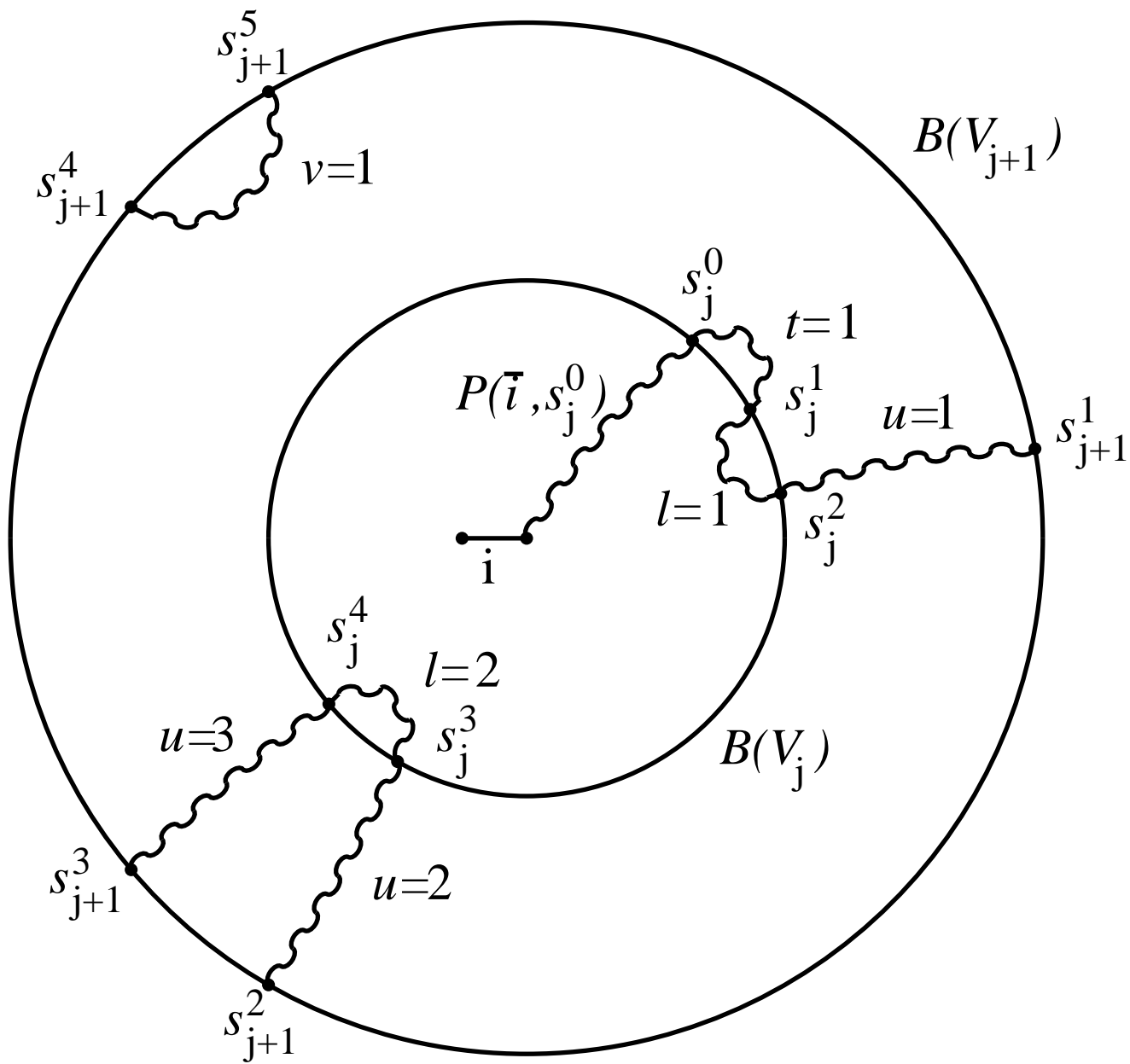
\item{\bf Fig. 1} We illustrate the case (see eq. (38)) where $m' = 3$, $m = 
3$,
i.e. there are 3 paths in $V_{j+1}$ and 3 paths in $V_j$ with $t_j = 1$, $u_j
= 3$ and $v_{j+1} = 1$. $B(V_{j+1})$ and $B(V_j)$ are respectively the 
borders of
$V_{j+1}$ and $V_j$.
\item{} The path $P(\bar{i}, s_j^0)$ relates one end of the propagator $i$ to
$s_j^0$ without going through $i$. Paths $P(s_j^{2\ell - 1} , s_j^{2 \ell})$
corresponding to $\ell = 1$, relating $s_j^1$ to $s_j^2$, and to $\ell = 2$,
relating $s_j^3$ to $s_j^4$ are shown, together with three paths $P(s_j^{
\ell u} ,
s_{j+1}^{\ell u})$ with $u = 1, 2, 3$ and one path $P(s_{j+1}^{2 \ell v - 2} 
,
s_{j+1}^{2 \ell v - 1})$ with $v = 1$.  
\vskip 3 mm

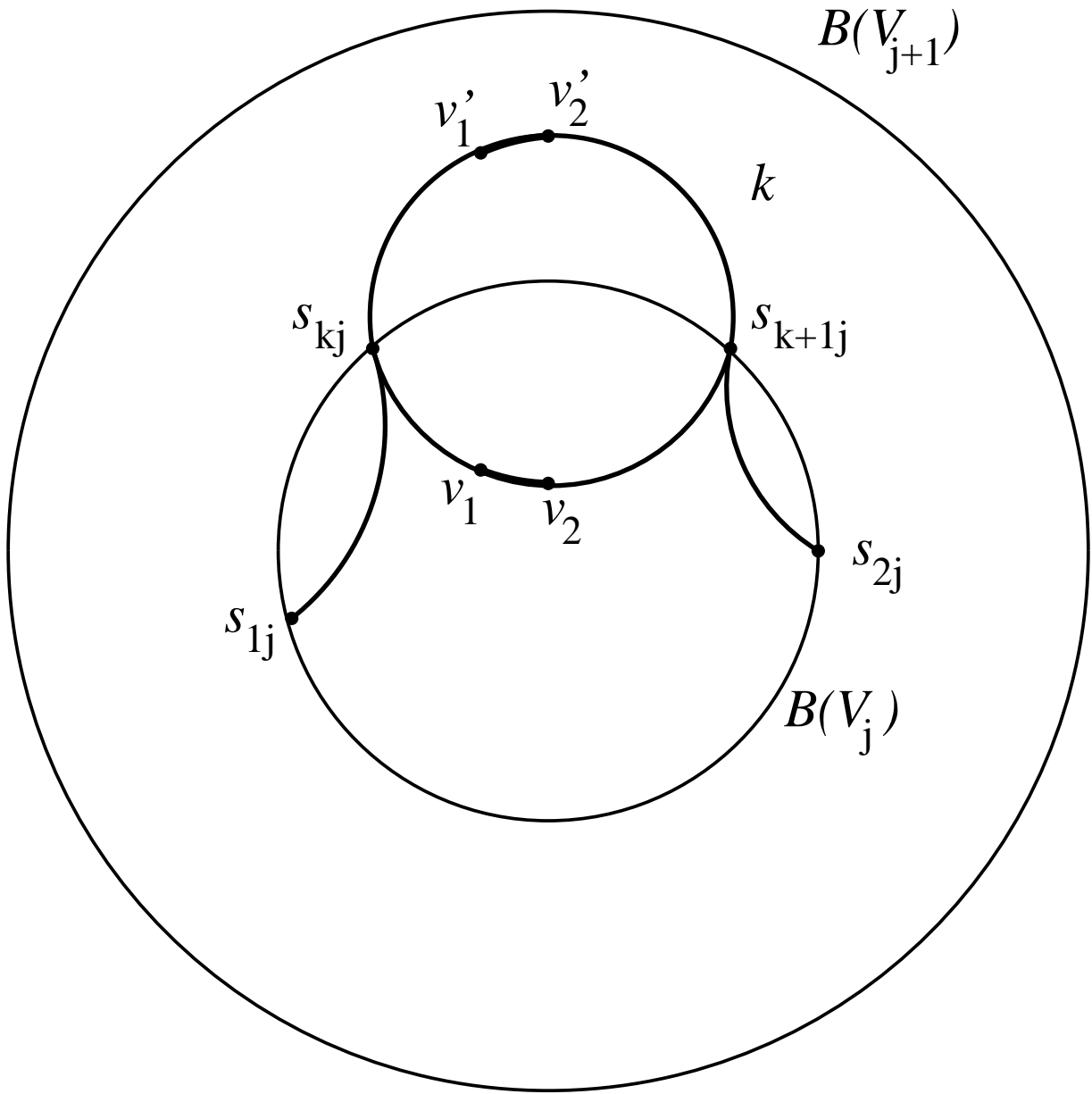
\item{\bf Fig. 2} The case where one loop ${\cal L}_k$ is present is 
depicted. The
path $P_1(s_{1j}, s_{2j})$ in $V_j$ consists of the successive paths 
$P(s_{1j},
s_{kj})$ $P_1(s_{kj}, s_{k+1j})$ $P(s_{k+1j}, s_{2j})$. The propagator
$(v_1v_2)$ is shown on $P_1(s_{kj}, s_{k+1j})$. The path $P_2(s_{1j}, 
s_{2j})$
differs from $P_1(s_{1j}, s_{2j})$ by the path $P_2(s_{kj}, s_{k+1j}$ in
$V_{j+1} - V_j$ on which the propagator $(v'_1 v'_2)$ is shown. The loop ${
\cal
L}_k$ is formed by the succession of paths $P_1(s_{kj}, s_{k+1j})$ $P_2(
s_{k+1j}, s_{kj})$ where the latter is the reverse path of $P_2(s_{kj},
s_{k+1j})$.
\vskip 3 mm 

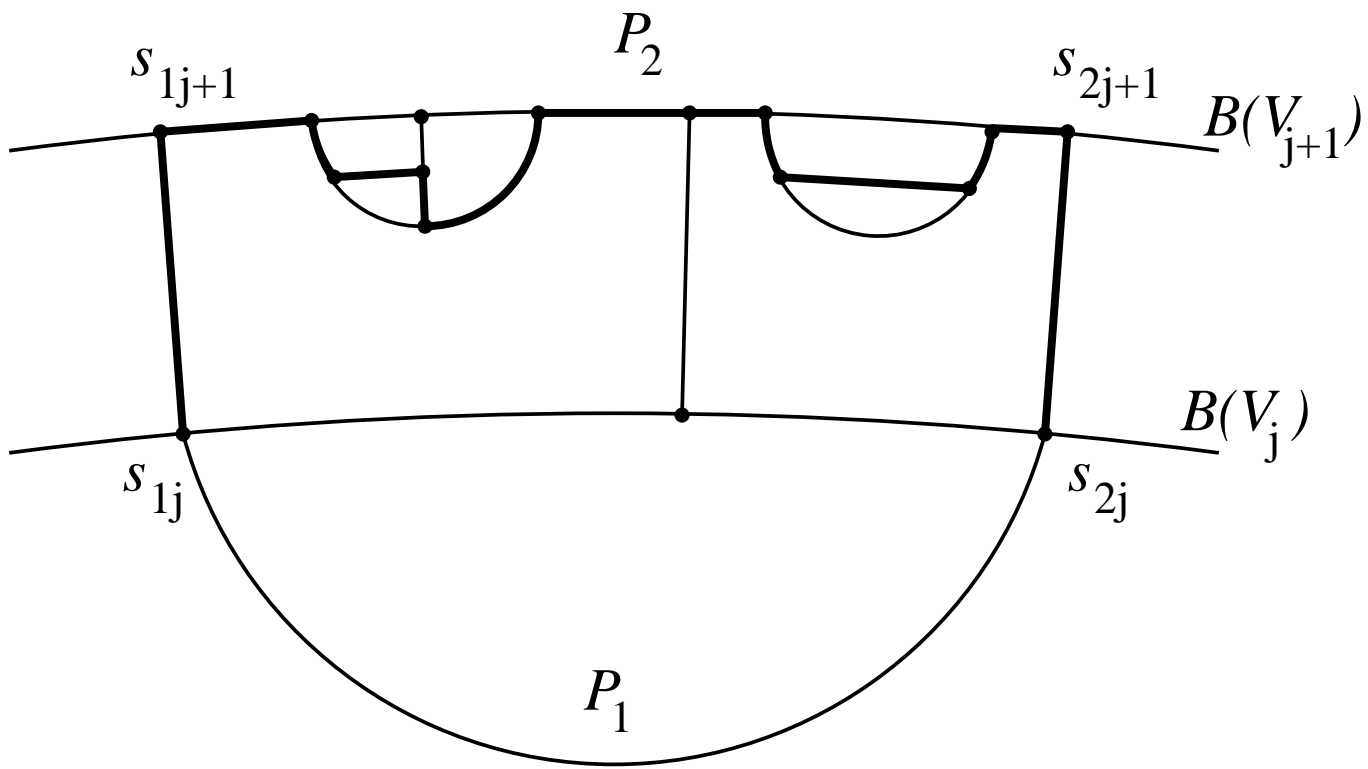
\item{\bf Fig. 3} The constraint that any propagator stemming out of $V_j$
should relate the part of the borders of $V_j$ and $V_{j+1}$ (on the deformed
spheres $S_{j+1}$ and $S_j$ respectively) has been imposed, if this 
propagator
is in $V_{j+1}$. The path in $V_j$ $P_1(s_{1j}, s_{2j})$ is shown (here ${
\cal
L}_k$ is simply ${\cal L}$). In $V_{j+1} - V_j$, the path $P_2(s_{1j},
s_{2j})$ is shown by a thick line. Depicted are the propagators $(s_{1j} \
s_{1j+1})$ and $(s_{2j} \ s_{2j+1})$. \par}

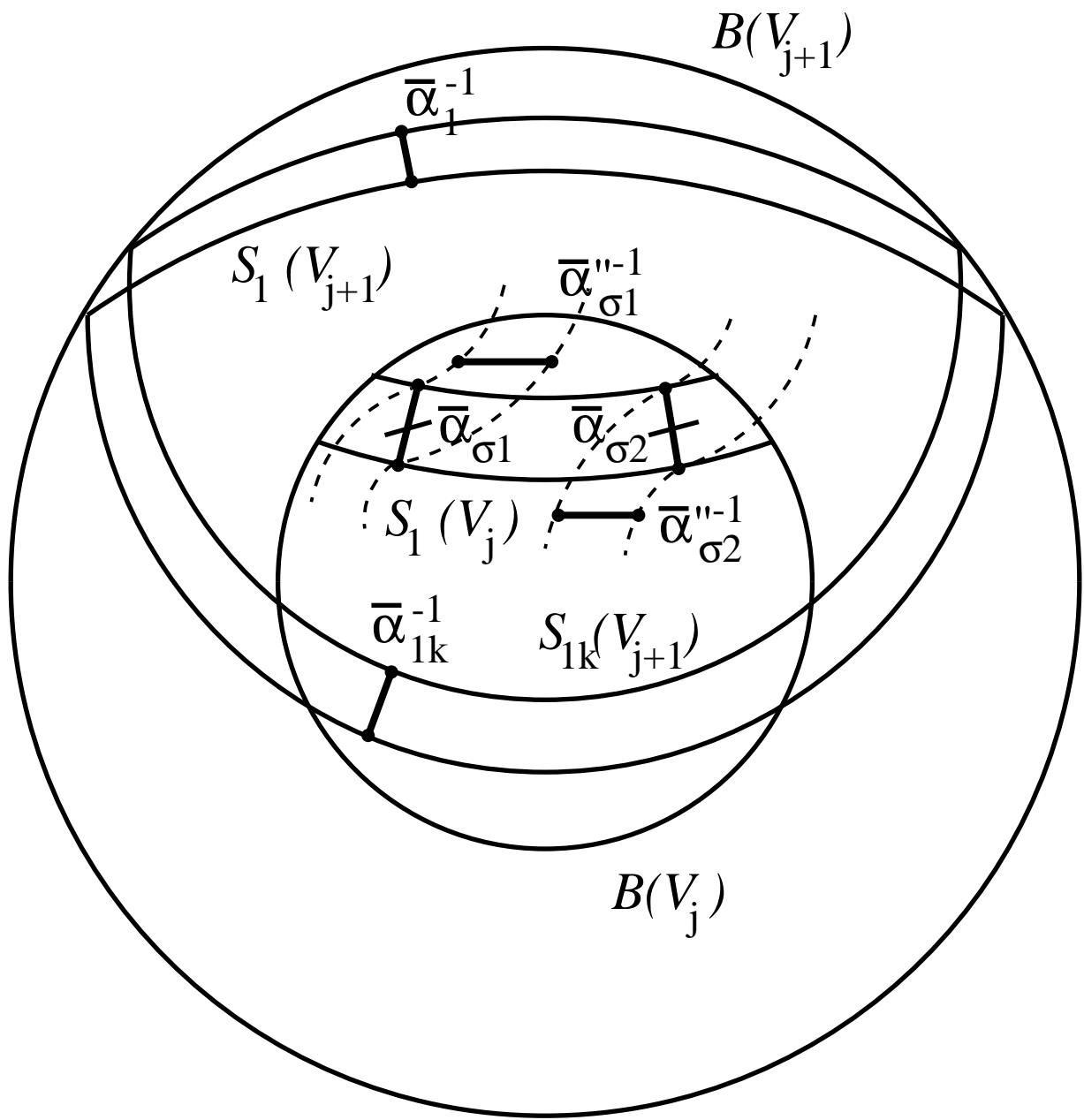
\bye